\documentclass[12pt]{article}

\usepackage{amsmath,graphicx}
\usepackage{epstopdf} 
\usepackage{multirow}
\usepackage{amsfonts}
\usepackage{amssymb}
\usepackage{amscd}

\def\hybrid{\topmargin -20pt    \oddsidemargin 0pt
        \headheight 0pt \headsep 0pt
        \textwidth 6.25in       
        \textheight 9.5in       
        \marginparwidth .875in
        \parskip 5pt plus 1pt   \jot = 1.5ex}

\hybrid

\numberwithin{equation}{section}
\numberwithin{table}{section}

\newcommand{\beq}{\begin{equation}\begin{aligned}}
\newcommand{\eeq}{\end{aligned}\end{equation}}
\newcommand{\bi}{\begin{itemize}}
\newcommand{\ei}{\end{itemize}}
\newcommand{\bea}{\begin{eqnarray}}
\newcommand{\eea}{\end{eqnarray}}
\newcommand{\ba}{\begin{array}}
\newcommand{\ea}{\end{array}}
\newcommand{\bt}{\begin{tabular}}
\newcommand{\et}{\end{tabular}}
\newcommand{\bc}{\begin{center}}
\newcommand{\ec}{\end{center}}

\newcommand{\cC}{C}

\newcommand{\cJ}{\mathcal{J}}

\newcommand{\cS}{\mathcal{S}}

\newcommand{\nn}{\nonumber}

\newcommand{\phiten}{\varphi}

\newcommand{\cref}{{\bf [check ref]}}

\newcommand{\phifour}{{\phi}}

\newcommand{\etac}{\eta}

\def\stt {$\SU(3) \times \SU(3)$}

\newcommand{\MQK}{{\mathcal M}_{\rm{QK}}}
\newcommand{\MHK}{{\mathcal M}_{\rm{HKC}}}
\newcommand{\MSK}{{\mathcal M}_{\rm{SK}}}
\newcommand{\Mrigid}{\widetilde{\mathcal{M}}_{\text{SK}}}
\newcommand{\MCY}{{\mathcal M}_{\rm{trunc}}}
\newcommand{\Mfull}{\widetilde{\mathcal{M}}}

\newcommand{\rePhi}{\rho}

\newcommand{\id}{\mathbf{1}}

\newcommand{\cP}{\mathcal{P}}

\newcommand{\dd}{\mathrm{d}}
\newcommand{\ee}{\mathrm{e}}
\newcommand{\ii}{\mathrm{i}}

\newcommand{\der}{\partial}

\newcommand{\bbZ}{\mathbb{Z}}
\newcommand{\bbR}{\mathbb{R}}
\newcommand{\bbC}{\mathbb{C}}

\DeclareMathOperator{\SL}{\mathit{SL}}
\DeclareMathOperator{\GL}{\mathit{GL}}
\DeclareMathOperator{\SU}{\mathit{SU}}
\DeclareMathOperator{\U}{\mathit{U}}
\DeclareMathOperator{\SO}{\mathit{SO}}
\DeclareMathOperator{\Oo}{\mathit{O}}
\DeclareMathOperator{\Symp}{\mathit{Sp}}
\DeclareMathOperator{\Spin}{\mathit{Spin}}
\DeclareMathOperator{\sll}{\mathfrak{sl}}
\DeclareMathOperator{\so}{\mathfrak{so}}
\DeclareMathOperator{\su}{\mathfrak{su}}

\DeclareMathOperator{\Es7}{\mathit{E}_{7(7)}}
\DeclareMathOperator{\es7}{\mathfrak{e}_{7(7)}}
\newcommand{\Tsub}{\SL(2,\bbR)\times O(6,6)}
\newcommand{\SLR}{\SL(2,\bbR)}
\DeclareMathOperator{\Ex6}{\mathit{E}_{6(2)}}

\newcommand{\rep}[1]{\mathbf{#1}}

\DeclareMathOperator{\tr}{tr}

\DeclareMathOperator{\re}{Re}
\DeclareMathOperator{\im}{Im}
\DeclareMathOperator{\vol}{vol}

\newcommand{\mukai}[2]{\big<{#1},{#2}\big>}

\newcommand{\tB}{\tilde{B}}
\newcommand{\tL}{\tilde{\Lambda}}

\newcommand{\Ea}{\mathcal{A}}
\newcommand{\Eb}{\mathcal{B}}
\newcommand{\Ec}{\mathcal{C}}
\newcommand{\Ed}{\mathcal{D}}

\newcommand{\Ta}{A}
\newcommand{\Tb}{B}

\newcommand{\JJ}{K}
\newcommand{\LL}{L}
\newcommand{\JSK}{J_\text{SK}}
\newcommand{\gSK}{g^\text{SK}}
\newcommand{\KSK}{K_\text{SK}}

\newcommand{\Jhit}{J_{\text{Hit}}}
\newcommand{\hg}{\SU(8)/\bbZ_2}

\newcommand{\smu}{\bar{\mu}}

\newcommand{\fb}{\nu}
\newcommand{\km}{\pi}

\newcommand{\hp}{\hphantom}

\begin{document}


\begin{titlepage}
\begin{center}


\rightline{\small IPhT-T09/039}
\rightline{\small ZMP-HH/09-3}
\rightline{\small Imperial/TP/2009/DW/01}
\vskip 1.4cm

{\Large \bf  $\Es7$ formulation of $N=2$ backgrounds}

\vskip 1.2cm

{\bf Mariana Gra{\~n}a$^{a}$, Jan Louis$^{b,c}$, Aaron Sim$^{d}$ and
Daniel Waldram$^{d,e}$ }

\vskip 0.4cm

{}$^{a}${\em Institut de Physique Th\'eorique,                   
CEA/ Saclay \\
91191 Gif-sur-Yvette Cedex, France}  \\
{\tt mariana.grana@cea.fr}

\vskip 0.4cm

{}$^{b}${\em II. Institut f{\"u}r Theoretische Physik der Universit{\"a}t Hamburg\\
Luruper Chaussee 149,  D-22761 Hamburg, Germany}\\
 {\tt jan.louis@desy.de} 

\vskip 0.4cm

{}$^{c}${\em Zentrum f\"ur Mathematische Physik, 
Universit\"at Hamburg,\\
Bundesstrasse 55, D-20146 Hamburg}

\vskip 0.4cm

{}$^{d}${\em Department of Physics, Imperial College London\\
London, SW7 2BZ, U.K.}\\
{\tt aaron.sim@imperial.ac.uk}, 
{\tt d.waldram@imperial.ac.uk}

\vskip 0.4cm

{}$^{e}${\em Institute for Mathematical Sciences, Imperial College London\\
London, SW7 2PG, U.K.}

\end{center}

\vskip 0.6cm

\begin{center} {\bf ABSTRACT } \end{center}

\noindent
In this paper we reformulate $N=2$ supergravity backgrounds arising in
type II string theory in terms of quantities transforming under the  U-duality 
group $\Es7$. In particular we combine the Ramond--Ramond scalar degrees of 
freedom together with the $\Oo(6,6)$ pure spinors which govern the
Neveu-Schwarz sector by considering an extended version of generalised
geometry. We give $\Es7$-invariant expressions for the K\"ahler and
hyperk\"ahler potentials describing the moduli space of vector and
hypermultiplets, demonstrating that both correspond to standard $\Es7$
coset spaces. We also find $\Es7$ expressions for the Killing
prepotentials defining the scalar potential, and discuss the equations
governing $N=1$ vacua in this formalism.  

\vfill

\noindent April 2009

\end{titlepage}



\section{ Introduction}


Backgrounds which involve manifolds with $G$-structure 
naturally appear in string theory as generalisations of Calabi--Yau
and other special holonomy compactifications~\cite{waldram,grana}. 
As for conventional special holonomy manifolds these backgrounds can
be classified by the amount of supersymmetry that they leave
unbroken. In both cases supersymmetry requires the existence of
nowhere vanishing and globally defined spinors. This in turn reduces
the structure group to a subgroup $G$ which leaves the spinors
invariant. 

For special holonomy manifolds the spinors are also covariantly
constant with respect to the Levi-Civita connection which is what 
in turn implies that the manifold has a reduced holonomy group. On the
other hand, the spinors of backgrounds with $G$-structure are
covariantly constant with respect to a different, torsionful
connection~\cite{waldram,Rocek,CS,salamonb}. In type II supergravity,
there are two spinors parameterising the supersymmetry. It is then
natural to consider a further generalisation to $G\times
G$-structures, with each spinor invariant under a different
$G$ subgroup. Geometrically this can be viewed~\cite{GMPT2} as a
structure on the sum of the tangent and cotangent spaces, using the
notion of ``generalised geometry'' first introduced by
Hitchin~\cite{Hitchin,Gualtieri}. In this case one can forget the
conventional geometrical structure on the manifold and discuss the
background just in terms of the $G\times G$-structures. It has the
advantage that these structures are often better defined globally and
also can satisfy integrability conditions that are the analogues of
special holonomy. 

From a particle physics point of view backgrounds which leave four
supercharges unbroken (corresponding to $N=1$ in four space-time
dimensions $(d=4)$) are the most interesting. However it is often
useful to first study backgrounds with additional supercharges as in
this case the couplings in the  effective action are more
constrained. In a series of papers~\cite{GLMW,GLW,GLW2} we considered
backgrounds with eight unbroken supercharges (corresponding to $N=2$
in $d=4$) and formulated them in the language of $SU(3)\times
SU(3)$-structures. 

In refs.~\cite{GLW,GLW2} we studied this problem from two different points
of view. On the one hand, by losing manifest $\SO(9,1)$-invariance one
can rewrite the ten-dimensional supergravity in a form where only
eight supercharges are manifest. This corresponds to a rewriting  
of the ten-dimensional action in ``$N=2$ form'' though without any
Kaluza-Klein reduction \cite{dWN}. A slightly different point of view
arises when one considers a Kaluza-Klein truncation keeping only the
light modes. In this case one can integrate over the six-dimensional
manifold and derive an ``honest'' $N=2$ effective action in $d=4$. In
this paper we will only consider the first approach.\footnote{Further
  aspects about both effective actions are discussed, for example, in
  refs.~\cite{Benmachiche:2006df}--\cite{Martucci:2009sf}.}

For $G\times G$-structures the ``unification'' of the tangent and
cotangent bundle suggests a formalism where instead of the
usual tangent space structure group $\GL(6,\bbR)$, the  group
$\Oo(6,6)$ is used, acting on the sum of tangent and cotangent spaces.
It turns out that the $N=2$ geometry is most naturally described by
two complex 32-dimensional (pure) spinors $\Phi^\pm$ of
$\Oo(6,6)$~\cite{Hitchin}. Each of them individually defines an 
$\SU(3,3)$ structure. The magnitude and phase of $\Phi^\pm$ are
unphysical, so each can be viewed as parameterising a point in an
$\Oo(6,6)$ orbit corresponding to the special K\"ahler coset space
${\cal M}_{\rm SK} = \Oo(6,6)/U(3,3)$. The respective K\"ahler
potentials can be expressed in terms of the square root of the 
$\Oo(6,6)$ quartic invariant built out of $\Phi^\pm$, known as the
Hitchin function. We review these results in detail in
section~\ref{O66}. 

The $\Oo(6,6)$ formalism naturally captures the degrees of freedom of
the NS-sector, i.e.\ the metric and the $B$-field, but it does not
incorporate the Ramond-Ramond (RR) sector four-dimensional scalars
into a geometrical description. One knows that for Calabi--Yau 
compactifications including the RR-scalars promotes the special
K\"ahler manifold ${\cal M}_{\rm SK}$ into a  dual 
quaternionic-K\"ahler (QK) space ${\cal M}_{\rm QK}$. 
The map ${\cal M}_{\rm  SK} \to {\cal M}_{\rm QK}$ is a generic
property  of type II string backgrounds and is called the
c-map~\cite{CFG,FS}. One can also consider the hyperk\"ahler cone (or the
Swann bundle) over ${\cal M}_{\rm QK}$~\cite{salamonswann,Swann}. Such
a construction always exists and physically corresponds to the 
coupling of hypermultiplets to superconformal supergravity
\cite{HKC,VPnotes}. The hyperk\"ahler cone has one extra quaternionic
dimension corresponding to a superconformal compensator multiplet.
The presence of the compensator gauges the $\SU(2)_{\rm R}$-symmetry
of $N=2$ together with a dilatation symmetry. The metric on the cone
is then determined by a hyperk\"ahler potential $\chi$.  

Thus the question arises if there is a generalisation of the
$\Oo(6,6)$ formalism which describes the deformation space ${\cal
  M}_{\rm QK}$. This is the topic of the present paper. 
  By analogy with the corresponding 
  discrete T- and U-duality groups, one wants to replace the  group $\Oo(6,6)$ of the NS-sector
by  $\Es7$ which acts non-trivially on all
scalar fields and mixes the scalars from the NS sector with the
scalars in the RR sector~\cite{HT}.\footnote{In this paper we will refer to these groups loosely as
T- and U-duality, though the connection to the actual discrete duality groups is only
clear for toroidal compactifications.}
 Geometrically, this ``extends''  
Hitchin's generalised geometry and includes the RR degrees of freedom
in a yet larger structure called ``extended geometry'' or
``exceptional generalised'' geometry
(EGG)~\cite{Chris,EGG}. 
It is important to note that $\Es7$ is not a symmetry of EGG (nor is
$\Oo(d,d)$ a symmetry of generalised geometry).\footnote{This is in
  contrast to more ambitious proposals such as~\cite{E11}-\cite{Hillmann}.}
Instead, the construction is covariant with respect to a subgroup
built from the diffeomorphism symmetry and the gauge transformations
of the NS and RR form-fields, and, in addition, the objects of
interest come naturally in $\Es7$ representations. 

We find that the quaternionic-K\"ahler manifold ${\cal M}_{\rm QK}$ 
is one of the homogeneous Wolf spaces~\cite{wolf,alex}, namely ${\cal
  M}_{\rm QK}= \Es7/(SO^*(12) \times \SU(2))$, for which the
hyperk\"ahler cone is ${\cal M}_{\rm HKC} = \bbR^+\times\Es7/\SO^*(12)$.
The latter space can be viewed as the moduli space of highest weight
$\SU(2)$ embeddings into $\Es7$~\cite{wolf}. 
{}From this construction a hyperk\"ahler potential $\chi$
can be given in terms of the $\SU(2)$
generators~\cite{KS-homo2,KS-Wolf}. By decomposing $\Es7$ under its
subgroup $\Tsub$ we specify explicitly the embeddings of one pure
spinor, the RR potential $C$ and the dilaton-axion. Then using
the result of~\cite{KS-Wolf} we are able to establish agreement with
the expression for $\chi$ given in \cite{RVV,pioline} for
hyperk\"ahler cones which generically arise via the c-map.   

We also find that the space $\Oo(6,6)/U(3,3)$ can be promoted to the
special $\Es7$ K\"ahler coset ${\cal M}_{\rm SK}=\bbR^+\times\Es7/\Ex6$
which again admits an action of the U-duality group $\Es7$. Furthermore
its K\"ahler potential is given by the square root of the $\Es7$
quartic invariant built out of the $\rep{56}$ representation. This
expression can be viewed as an $\Es7$ Hitchin function.

The $\Es7$ cosets just discussed do not appear directly in the $N=2$
supergravity but a compatibility condition between the two spinors
$\Phi^\pm$ (or more precisely, between the $SO^*(12)$ and $\Ex6$
structures) has to be imposed. Furthermore, if the low-energy theory
is to contain no additional massive gravitino multiplets, they
either have to be integrated out or an appropriate projection
is required. As these massive $N=2$ gravitino multiplets
contain scalar degrees of freedom, the scalar geometry is altered.
This is reviewed in more detail in section~\ref{O66}.

In addition to the kinetic terms, the scalar potential can also be
expressed in an $\Es7$ language, though now in a way that depends on the 
differential geometry of the EGG. Generically in $N=2$ supergravity the scalar
potential is given in terms of an $\SU(2)$-triplet of Killing
prepotentials $\cP_a$. We propose an $\Es7$ form for 
$\cP_a$ which coincides with the known expressions given in
\cite{GLW,GLW2} when decomposed under the $\Tsub$ subgroup of
$\Es7$. One can also consider the form of the $N=1$ vacuum equations
in this formalism. We do not give a complete description here but show
at least how the standard $O(6,6)$ equations~\cite{GMPT2} can be embedded as particular components of $\Es7$ expressions. 

This paper is organised as follows. Throughout, for definiteness, we
focus on the case of type IIA backgrounds, though the same formalism
works equally well for type IIB. In section \ref{O66}
we recall how the $N=2$ backgrounds can be written in terms of the
generalised geometrical $\Oo(6,6)$ 
formalism following \cite{GLW,GLW2}. In section \ref{reform} we include the RR
degrees of freedom and formulate the combined structure in terms of an 
exceptional generalised geometry (EGG)~\cite{Chris,EGG}. In particular, in section 
\ref{app:Tsub} we first give some basic $\Es7$ definitions and in 
\ref{EGG} we introduce the notion of exceptional
generalised geometry. Then in sections \ref{hyper} and \ref{vectors}
we discuss the moduli spaces for the hypermultiplet 
and vector multiplet sectors in terms of $\Es7$ coset manifolds,
specifying in particular  the $\Tsub$ embedding of the NS and RR degrees of freedom.
We give an $\Es7$-invariant expression for the hyperk\"ahler potential $\chi$ following the explicit construction of 
ref.~\cite{KS-Wolf} and we also show that the K\"ahler potential on the vector multiplet 
moduli space ${\cal M}_{\rm SK}^+$ is given by the
square-root of the $\Es7$ quartic invariant in complete analogy to the
Hitchin function in the $O(6,6)$ case. In section \ref{compat}
we discuss the combined vector and hypermultiplet sectors, 
which results in some compatibility condition between the structures in $\Es7$,
as well as some constraints coming from requiring a standard supergravity action.  
In section \ref{E7P} we then give an
$\Es7$ expression for the Killing prepotentials.
Section \ref{N=1text} discusses the $\Es7$ version of the 
$N=1$ background conditions determined in ref.~\cite{GMPT2}.
Section~\ref{Conc} contains our conclusions and some of the more technical details of the computations are presented in two appendices.


\section{Review of $\Oo(6,6)$ formalism}
\label{O66}

In this section we briefly recall some of the results of
refs.~\cite{GLW,GLW2} in order to set the stage for our analysis. We will use the conventions
of \cite{GLW2}. It
is perhaps helpful to stress again 
that in this paper we are \emph{not} making a
dimensional reduction of type II supergravity. Rather we are rewriting
the full ten-dimensional theory in a four-dimensional $N=2$ language,
where one can decompose the degrees of freedom into hypermultiplets
and vector multiplets. Necessarily this requires breaking the manifest
local $\Spin(9,1)$ symmetry to $\Spin(3,1)\times\Spin(6)$, and also
that we can consistently pick out eight of the 32 supersymmetries. One
can then introduce special K\"ahler and quaternionic moduli spaces for
the corresponding scalar (with respect to $\Spin(3,1)$) degrees of
freedom. However these degrees of freedom will still depend on the
coordinates of all ten dimensions. 

As an example, suppose we have a product manifold
$M^{9,1}=M^{3,1}\times M^6$ with an $\SU(3)$ structure on $M^6$
defined by a two-form $J$ and a three-form $\Omega$, both of which are
scalars with respect to  $\Spin(3,1)$. If we had a Calabi--Yau
manifold then $\Omega$ and $J$ are constrained by requiring $\dd
J=\dd\Omega=0$. Let us focus on $\Omega$. In a conventional
dimensional reduction one expands  $\Omega$ in terms of harmonic forms
$(\alpha_A, \beta^A)$ according to
$\Omega=Z^A\alpha_A - F_B(Z)\beta^B$, and then shows 
that there is a special K\"ahler moduli space $\MCY$ for the
four-dimensional fields $Z^A$ which depends on the complex geometry of
the Calabi--Yau manifold. Similarly, for manifolds of $SU(3)$ structure, where 
nowhere vanishing $J$ and $\Omega$ exist but are generically not closed, one can 
truncate the degrees of freedom to a finite dimensional subspace,
and do a similar expansion as in Calabi-Yau manifolds, but in this case involving forms which are
not necessarily harmonic. The moduli space $\MCY$ spanned by the four-dimensional fields is still
special K\"ahler \cite{GLW2,mkp}. 
In this paper on the other hand we look at the space of all
structures $\Omega$. Rather than a finite set of moduli $Z^A$ one can choose a
different three-form $\Omega$ at each point in the six-dimensional
space. The space of such $\Omega$ at a given point is
$\bbR^+\times\GL(6,\bbR)/\SL(3,\bbC)$ and it turns out that this is
\emph{also} a special K\"ahler space $\MSK$. In summary, we have two
cases, given  $x\in M^{3,1}$  and $y\in M^6$ 
\begin{equation}
\begin{aligned}
   &\text{untruncated:} & \Omega &= \Omega(x,y) \in
   \Lambda^3TM^6 \ ,\\
   && \Omega(x,y) &\in \MSK 
   \qquad \text{at each point $(x,y)\in M^{3,1} \times M^6$ } \ ,\\
   &\text{finite truncation:} & 
      \Omega &= Z^A(x)\,\alpha_A(y) - F_B(Z(x))\,\beta^B(y) \ ,\\
   && Z^A(x) &\in \MCY
   \qquad \text{at each point $x\in M^{3,1}$} \ .\\
\end{aligned}
\end{equation}
Note that $\MSK\simeq\bbR^+\times\GL(6,\bbR)/\SL(3,\bbC)$ is the same for
all manifolds $M^6$ while $\MCY$ depends on the particular
manifold. Furthermore $\MCY$ can be obtained from the fibration of
$\MSK$ over $M^6$ by restricting to a finite subspace of sections $\Omega$. 

More generally in~\cite{GLW,GLW2} we simply assumed that the
tangent bundle of the ten-dimensional space-time  splits according to
$TM^{9,1}=T^{3,1}\oplus F$, where $F$ admits a pair of nowhere
vanishing $\Spin(6)$-spinors. Here, for simplicity, we will 
always consider the case where $M^{9,1}=M^{3,1}\times M^{6}$ so
$F=TM^6$ and usually just write $TM$ for $TM^6$. The split of the
tangent space implies that all fields of the theory can be decomposed
under $\Spin(3,1)\times\Spin(6)$. In particular one can decompose the
two supersymmetry parameters of type II supergravity
$\epsilon^1,\epsilon^2$ as\footnote{In section \ref{compat} we will find some
subtleties in counting the degrees of freedom on the moduli spaces, that
actually will lead us to slightly generalise this $N=2$ spinor
ansatz.} 
\begin{equation}
\begin{aligned}
\label{decompepsilon}
   \epsilon^1 &= \varepsilon_+^1 \otimes \eta^1_-
      + \varepsilon_-^1 \otimes \eta^1_+ \ , \\
   \epsilon^2 &= \varepsilon_+^2 \otimes \eta^2_\pm 
      + \varepsilon_-^2 \otimes \eta^2_\mp \ ,
\end{aligned}
\end{equation}
where in the second line the upper sign is taken for type IIA and the
lower for type IIB. Here $\eta^I_+$ with $I=1,2$ are 
spinors of $\Spin(6)$ while $\varepsilon^I$ are Weyl spinors of
$\Spin(3,1)$.\footnote{In each case
$\eta^I_-$ and $\varepsilon^I_-$ are the charge conjugate spinors and
the $\pm$ subscripts denote the chirality (for more details see
appendix~A of \cite{GLW2}).} We see that for a given pair
$(\eta_+^1,\eta_+^2)$ there are eight spinors parameterised by
$\varepsilon^I_\pm$.  
These are the eight supersymmetries which remain
manifest in the reformulated theory. 
Each of the $\eta^I$ is invariant under a (different) $\SU(3)$ inside
$\Spin(6)$. The two $\SU(3)$ intersect in an $\SU(2)$ and the
established nomenclature calls this situation a local
$\SU(2)$-structure. 

Such backgrounds have a very natural interpretation in terms of
generalised geometry. Recall that this is defined in terms of the
generalised tangent space 
\begin{equation}
   E=TM\oplus T^*M
\end{equation}
built from the sum of the tangent and cotangent spaces. If $M$ is
$d$-dimensional, there is a natural $\Oo(d,d)$-invariant
metric\footnote{We use $\eta$ to denote both the $O(d,d)$ metric
  and the $O(d)$ spinors $\eta^I$. The distinction between
  them should be clear from the context.}
on $E$, given by $\etac(Y,Y) = i_y \xi$ where $Y=y+\xi\in E$, with $y\in
TM$ and $\xi\in T^*M$. One can then combine $(\eta^1,\eta^2)$ into two
32-dimensional complex ``pure'' spinors $\Phi^\pm\in S^\pm$ of
$\Oo(6,6)$. They are defined as the spinor bilinears, or equivalently
sums of odd or even forms,  
\begin{equation}
\label{purespinors}
   \Phi^+ = \ee^{-B} \eta^1_+\bar{\eta}^2_+ \equiv \ee^{-B} \Phi_0^+
     \ , \qquad
   \Phi^- =  \ee^{-B} \eta^1_+\bar{\eta}^2_- \equiv \ee^{-B} \Phi_0^- ,
\end{equation}
In the special case where the two spinors are aligned we have
$\eta^1=\eta^2\equiv\eta$. In this case there is only  a single $\SU(3)$
structure, familiar from the case of Calabi--Yau compactification, and
one has
\beq\label{SU3limit}
\Phi^+= \ee^{-(B+\ii J)}\ ,\qquad 
\Phi^-=- \ii \ee^{-B} \Omega\ ,
\eeq
where $\Omega$ is the complex $(3,0)$-form and $J$ is the real $(1,1)$-form.

Each pure spinor is invariant under an $\SU(3,3)$ 
subgroup of $\Oo(6,6)$ and so each individually is said to define an
$\SU(3,3)$ structure on $E$. In particular this defines a 
generalised (almost) complex structure. Explicitly one can construct the
invariant tensor  
\begin{equation}
\label{Jgen}
   \mathcal{J}^{\pm\Ta}{}_\Tb\ =\ \ii\, \frac{
       \mukai{\Phi^\pm}{\Gamma^\Ta{}_\Tb\bar{\Phi}^\pm}}
       {\mukai{\Phi^\pm}{\bar{\Phi}^\pm}}\ ,
\end{equation}
satisfying $(\mathcal{J}^\pm)^2=-\id$. Here, $\Gamma^\Ta$ with
$\Ta=1,\dots,12$ are gamma-matrices of $O(6,6)$, $\Gamma^{\Ta\Tb}$ are
antisymmetrised products of gamma-matrices, indices are raised and
lowered using $\etac$ and the bracket denotes the Mukai
pairing defined by   
\begin{equation}
\label{mukai}
   \mukai{\psi}{\chi} 
     = \sum_p (-)^{[(p+1)/2]} \psi_p \wedge \chi_{6-p} 
     \equiv (s (\psi) \wedge \chi)_6 \ .
\end{equation}
(The subscripts denote the degree of the component forms, and the
operation $s$ assigns the appropriate signs to the component forms. This
pairing is simply the natural real bilinear on $O(6,6)$ spinors. Note
that the pure spinors $\Phi^\pm$ also satisfy
$\mukai{\Phi^+}{\bar{\Phi}^+}=\mukai{\Phi^-}{\bar{\Phi}^-}$.)  

The generalised almost complex structures $\mathcal{J}^\pm$ also
induce a decomposition of the generalised spinor bundles $S^\pm$ into 
modules with definite eigenvalue under the action
of $\frac{1}{4}\mathcal{J}^\pm_{\Ta\Tb}\Gamma^{\Ta\Tb}$. In particular,
one finds
\begin{equation}
   \tfrac14 \cJ^\pm_{AB} \Gamma^{AB}\, \Phi^\pm = 3\ii \Phi^\pm, 
   \qquad 
   \tfrac14 \cJ^\pm_{AB} \Gamma^{AB}\, \bar\Phi^\pm 
      = -3\ii \bar\Phi^\pm\ .
\end{equation}
One can also use this action to define a coarser grading of $S^\pm$,
namely an almost complex structure on $S^\pm$, first introduced in
this context by Hitchin~\cite{Hitchin}, and given by
\begin{equation}
\label{Jhit}
   \Jhit^\pm = \exp\left( 
      \tfrac{1}{8}\pi\mathcal{J}_{\Ta\Tb}^\pm\Gamma^{\Ta\Tb}\right) \ ,
\end{equation}
such that (in six-dimensions and acting on $S^\pm(E)$) one has
$(\Jhit^\pm)^2=-\id$ and in particular,
$\Jhit^\pm\Phi^\pm=-\ii\Phi^\pm$.  

One finds that the specific $\Phi^\pm$ given by \eqref{purespinors}
also satisfy the ``compatibility'' condition 
\begin{eqnarray}\label{comp1}
   \mukai{\Phi^+}{ \Gamma^\Ta\Phi^-} = 0  \quad \forall A\ .
\end{eqnarray}
This implies that the common stabiliser group in $O(6,6)$ of the pair
$(\Phi^+,\Phi^-)$ is $\SU(3)\times\SU(3)$, or equivalently that
together they define an $\SU(3)\times\SU(3)$ structure in $O(6,6)$. 

One can also view this structure in terms of the way the supergravity
metric $g$ and $B$-field are encoded in generalised geometry. One can
combine $g$ and $B$ into an $O(2d)$ metric on the generalised tangent
space. This is compatible with the $O(d,d)$ metric such that together they
are invariant under $O(d)\times O(d)$ and hence define an $O(d)\times
O(d)$ structure. Thus in the six-dimensional case one can regard $g$
and $B$ as parameterising the 36-dimensional (Narain) coset space  
$O(6,6)/O(6) \times O(6)$. The two six-dimensional spinors $\eta^I$
transform separately under the two $\Spin(6)$ groups. Therefore the nowhere
vanishing pair $(\eta^1,\eta^2)$ defines a separate $\SU(3)$
structure in each $O(6)$ factor. Thus collectively we see that $g$,
$B$ and the pair $(\eta^1,\eta^2)$ define an $\SU(3)\times\SU(3)$
structure in generalised geometry. 

In summary we conclude that each pure spinor $\Phi^\pm$ defines an
$\SU(3,3)$ structure and that each parameterises a 32-dimensional coset
space \cite{Hitchin}
\begin{equation}
\label{rigid}
   \Mrigid^\pm 
      = \frac{\Oo(6,6)}{\SU(3,3)} \times \bbR^+ . 
\end{equation}
The appearance of the coset $G/H$ can also be understood as
follows. It is the orbit generated by the $G$-action on an element
which is stabilised by $H$. A simple example is the sphere
$S^d=SO(d+1)/SO(d)$, which can be seen as the orbit of the unit vector
in $\Bbb R^{d+1}$ when acting with the group $SO(d+1)$. We have
precisely the same situation in that $\Phi^+$, say, can be viewed as
parameterising ${\Oo(6,6)}$ orbits which are stabilised by
${\SU(3,3)}$. The $\bbR^+$ then corresponds to the freedom to
additionally rescale $\Phi^+$. In fact the real part of $\Phi^\pm$
alone is stabilised by $\SU(3,3)$. Since a generic real spinor is 32
dimensional, as are the orbits, we see that in this case the orbit of
$\re\Phi^\pm$ forms an open set in the space of all real spinors (a so
called ``stable orbit'')~\cite{Hitchin}. 

It turns out that the magnitude and phase of
$\Phi^\pm$ are not physical. Modding out by such complex rescalings
gives the spaces 
\begin{equation}
\label{MSK}
   \mathcal{M}^\pm_{\text{SK}} = \Mrigid^\pm/\bbC^*
      \simeq  \frac{\Oo(6,6)}{U(3,3)} \ . 
\end{equation}
As we will review below, there is a natural rigid special K\"ahler
metric on $\Mrigid^\pm$ and a local special K\"ahler metric on
$\mathcal{M}_{\text{SK}}^\pm$. The K\"ahler
potentials read \cite{Hitchin}
\beq \label{Kahlerpot}
\ee^{-K^\pm} 
   = \ii  \mukai{\Phi^\pm}{\bar \Phi^\pm} \ .
\eeq
Note that a
complex rescaling of the pure spinors $\Phi^\pm$ is unphysical in that
it corresponds to a K\"ahler transformation in $K^\pm$. This degree of
freedom in $\Phi^\pm$ will be part of a superconformal compensator in
the $\Es7$ formulation. 

Given that the groups $\SO(6,6)$ and
$\SU(3,3)$ are non-compact, the spaces $\MSK^\pm$ and $\Mrigid^\pm$
are both non-compact and have pseudo-Riemannian metrics on them. In
particular the signature of the metric on $\MSK^\pm$ is $(18,12)$. We
return to this below. 

As we have mentioned above, the two $\re \Phi^\pm$ together satisfying
\eqref{comp1} define an $\SU(3) \times \SU(3)$ structure inside
$\Oo(6,6)$. Therefore the compatible pair $(\re \Phi^+, \re \Phi^-)$
parameterises the 52-dimensional coset 
\begin{equation}
   (\re \Phi^+, \re \Phi^-) : \qquad 
   \Mfull = \frac{\Oo(6,6)}{\SU(3) \times \SU(3)} 
       \times  \bbR^+ \times \bbR^+ \ .
\end{equation}
(Note that the dimensionality of $\Mfull$ counts correctly the $2 \times 32$
degrees of freedom in $\re \Phi^+, \re \Phi^-$ minus the 12
compatibility constraints of \eqref{comp1}.)  $\Mfull$ is a particular
slice in the product space $\MSK^+\times\MSK^-$. Again for the physical
moduli space one needs to mod out by the $\bbC^*$ actions on $\Phi^\pm$, giving
the 48-dimensional coset $O(6,6)/U(3)\times U(3)$. 

Note, however, that this counting still does not match the physical NS
supergravity degrees of freedom which is the 36-dimensional space of
$g$ and $B$, parameterising the Narain coset
$O(6,6)/O(6)\times O(6)$. Furthermore, we note that the metric
on $O(6,6)/U(3)\times U(3)$ has signature $(36,12)$. Thus there are
twelve degrees of freedom in the latter coset which are not really
physical (and have the wrong sign kinetic term). Under $SU(3)\times
SU(3)$ these transform as triplets $(\rep{3},\rep{1})$,
$(\rep{1},\rep{3})$ and their complex conjugates. In terms of $N=2$
supergravity, these representations are associated with the massive
spin-$\frac32$ multiplets and one expects
that these directions are gauge degrees of freedom of the 
massive spin-$\frac32$ multiplets.
This leaves a 36-dimensional space as the physical parameter
space.  It would be interesting to give a geometrical
interpretation of this reduction, perhaps as a symplectic
reduction of $\MSK^+\times\MSK^-$ with a moment map corresponding to
the constraint~\eqref{comp1}.

We can make this physical content explicit by using the decomposition 
under $\SU(3)\times\SU(3)$ to assign the deformations along the orbits
of $\Phi^+$ and $\Phi^-$ as well as the RR degrees of freedom to $N=2$ multiplets.
In type IIA, the RR potential contains forms of odd degree, which from
the four-dimensional  point of view
contribute to vectors and scalars. The vectors, having one space-time index, are even forms 
on the internal space and we denote them $C_{\mu}^+$, while the scalars are internal odd forms denoted $C^-$. 
In order to recover the
standard $N=2$ supergravity structure we imposed in
refs.~\cite{GLW,GLW2} the constraint that no massive
spin-$\frac32$ multiplets appear. As we mentioned, this corresponds to
projecting out any triplet of the form $\rep{(3,1)}, \rep{(1,3)}$ or
their complex conjugates. With this projection only the gravitational
multiplet together with hyper-, tensor-, and vector multiplets survive.
These are shown for type IIA in Table~\ref{N=2multipletsA}. 
(In what follows, we restrict to type IIA theory,
the type IIB case follows easily by changing chiralities.)
\begin{table}[h]
\begin{center}
\begin{tabular}{|c|c|c|} \hline
 \rule[-0.3cm]{0cm}{0.8cm}
multiplet & \stt rep. & bosonic field content\\ \hline
\rule[-0.3cm]{0cm}{0.8cm}
 gravity multiplet&$(\rep 1, \rep{1}) $ & $ g_{\mu \nu}, \cC^{+}_{\mu \, (\rep{1})}$ \\ \hline
\rule[-0.3cm]{0cm}{0.8cm}
 vector multiplets& $(\rep{3}, \rep{\bar 3})$ &  $\cC^+_{\mu},  \Phi^+
$\\ \hline
\rule[-0.3cm]{0cm}{0.8cm}
{hypermultiplets } & $(\rep{3}, \rep{3})$&   $  \Phi^-, \cC^- $ \\
\hline
\rule[-0.3cm]{0cm}{0.8cm} 
{tensor multiplet} & $(\rep 1, \rep 1)$& $ B_{\mu\nu}, \phifour, \cC^{-}_{(\rep{1})} $ \\ \hline
\end{tabular}
\caption{\small 
\textit{N=2 multiplets in type IIA}}\label{N=2multipletsA}
\end{center}
\end{table}
$g_{\mu \nu}$ and  
$C_{\mu \, (\rep{1})}^+$ denote the graviton and the graviphoton, respectively,
which together form the bosonic components of the gravitational 
multiplet.\footnote{The
subscript  $(\rep{1})$ indicates that it is the $\SU(3)\times\SU(3)$ 
singlet of the RR forms $C_{\mu}^+$ or $C^-$.}
$\Phi^+$ represents the scalar degrees of freedom in the vector multiplets
(with the $(\rep{3,\bar 3})$ part of $C_\mu^+$ being the vectors)
while $\Phi^-$ together with $C^-$ combines into a hypermultiplet.
Finally the four-dimensional dilaton $\phifour$, $B_{\mu\nu}$ and 
the \stt\ singlet component of $C^-_{(\rep{1})}$ form 
the universal tensor multiplet.

After requiring compatible spinors and projecting out the triplets,
$N=2$ supergravity requires the scalar manifold to be
\begin{equation}
 \label{mod}
 {\cal M} = {\cal M}_{\rm SK}^+ \times  {\cal M}_{\rm QK}\ ,
 \end{equation}
where the first factor arises from $\Phi^+$ (or more generally from
the vector multiplets), while the second factor comes from 
the hypermultiplets $(\Phi^-,C^-)$ and the dualised tensor multiplet.
In the NS-subsector, i.e.~for
$C^-=0$, one has the submanifold 
\begin{equation}
 \label{modNS}
 {\cal M}_{\textrm{NS}} = {\cal M}_{\rm SK}^+ 
 \times  {\cal M}_{\rm SK}^- \times \frac{SU(1,1)}{U(1)}\ ,
 \end{equation}
where the K\"ahler
potentials of the first two factors are still given by
(\ref{Kahlerpot})  \cite{Strominger,GLMW,GLW,GLW2}, 
while for the last factor it reads
\beq \label{Kahlerpotdil}
\ee^{-K_S} =-\ii (S-\bar S) = 2\ee^{-2\phi} \ .
\eeq
 The four-dimensional dilaton $\phi$ is related to the
ten-dimensional dilaton $\phiten$ by $\phi=\phiten-\frac{1}{4}\ln\det
g_{mn}$. Equivalently one can write 
\begin{equation}
\label{eq:phifour}
   \ee^{-2\phi} = \ee^{-2\phiten}\vol_6 \ , 
\end{equation}
where $\vol_6$ is the volume form on $M$, so $\ee^{-2\phi}$ transforms
as a six-form. In $S$ it combines with the six-form $\tilde B$
corresponding to the ten-dimensional dual of $B_{\mu\nu}$, into the
complex  six-form field $S=\tilde B+\ii\ee^{-2\phi}$. 
  Let us stress that, even though we are using the same
notation, the individual factors in \eqref{modNS} are \emph{not} given
by \eqref{MSK}.  The latter only appear before applying the
triplet-projection and the compatibility constraint. 

In the case of a single $SU(3)$ structure the K\"ahler potentials
(\ref{Kahlerpot}) reduce to the familiar Calabi-Yau 
expressions \cite{Strominger}. Inserting \eqref{SU3limit} into
\eqref{Kahlerpot} one arrives at 
\beq \label{Kahlerpotsingle}
   \ee^{-K^+}=\tfrac{4}{3} J \wedge J \wedge J\ ,  \qquad
   \ee^{-K^-}= \ii \Omega \wedge \bar \Omega\ .
\eeq
The exponentials $\ee^{-K^\pm}$ in \eqref{Kahlerpot} coincide with the
Hitchin function $H$ defined for stable spinors of $O(6,6)$. If we
write $\rePhi^\pm=2\re\Phi^\pm$ then $H$ is the square root of the
spinor quartic invariant $q(\rePhi^\pm)$ of $O(6,6)$, that is 
\beq \label{KrigidH}
   \ee^{-K^\pm}
      = H(\rePhi^\pm)
      =\sqrt{q(\rePhi^\pm)}\ , 
\eeq
where
\begin{equation}
\label{Qdef}
   q(\rePhi) 
      = \tfrac{1}{48}\mukai{\rePhi}{\Gamma_{\Ta\Tb}\rePhi}
       \mukai{\rePhi}{\Gamma^{\Ta\Tb}\rePhi}  \ .   
\end{equation}
As was first shown by Hitchin \cite{Hitchin}, given that 
the Mukai pairing defines a
symplectic structure, the Hitchin function 
encodes the complex structure~\eqref{Jhit} such that together they
define a rigid special K\"ahler metric on $\Mrigid^\pm$ and hence a
local special K\"ahler metric on $\mathcal{M}_\text{SK}^\pm$. In
particular one can construct a second spinor $\hat{\rePhi}^\pm$ from
$\der H/ \der\rePhi^\pm$. Writing
$\Phi^\pm=\frac{1}{2}(\rePhi^\pm+\ii \hat{\rePhi}^\pm)$, the Hitchin function
is given by the expression (\ref{Kahlerpot}).    

To complete the description of the ten-dimensional supergravity in
terms of $N=2$ language, we must give the Killing prepotentials $\cP_a$ which determine the $N=2$ scalar potential. These are similarly
expressed in terms of $\Phi^\pm$ and can be written in a $O(6,6)$
form. For type IIA they read~\cite{GLW2}\footnote{Note that here
  we have made an $\SU(2)_R$ rotation
  $(\cP_1,\cP_2,\cP_3)\mapsto(\cP_1,-\cP_2,-\cP_3)$ as compared to
  the expressions in~\cite{GLW2}.} 
\begin{equation}
\label{Pdef}
\begin{aligned}
\cP_+ = \cP_1 +\ii\cP_2 
   &= -2\ii \ee^{\frac12 K^-+\phifour}\mukai{\Phi^+}{\dd\Phi^-} \ , \\
\cP_3 &= -\frac{1}{\sqrt{2}}\ii \ee^{2\phifour} \mukai{\Phi^+}{G}  \ .
\end{aligned}
\end{equation}
Note here we have introduced the closed RR field strengths
$G=\ee^{-B}F$ where $F$ are the more conventional field strengths
satisfying $\dd F-H\wedge F=0$. It will be useful to introduce a
potential for $G$, somewhat unconventionally denoted\footnote{This potential is usually called $A$ in the
  literature.} $C^-$, 
\beq
\label{GCdef}
   G=\ee^{-B} F\equiv \sqrt{2}\dd C^- \ , 
\eeq
where the factor of $\sqrt{2}$ is introduced to match the $\Es7$
conventions in what follows.


\section{Reformulation in terms of $\Es7$ and EGG}
\label{reform}


In this section we are extending the formalism reviewed in the
previous one by including the RR degrees of freedom $C^-$ (for
definiteness, we will consider the case of type IIA). Intuitively this
extension can be understood as promoting the 
T-duality group $\Oo(6,6)$ to the full U-duality group $\Es7$ which
acts on all degrees of freedom (not only the ones in the NS-sector)
and in particular mixes NS with RR scalars. From the supergravity
point of view adding RR scalars promotes the moduli space ${\cal
  M}_{\textrm{NS}}$ given in \eqref{modNS} to the moduli space ${\cal
  M} ={\cal M}_{\rm SK}^+ \times  {\cal M}_{\rm QK}$ given in
\eqref{mod}.  In particular one of special K\"ahler manifolds (${\cal
  M}_{\rm SK}^-$ for type IIA) together with the dilaton factor is
enlarged to a quaternionic-K\"ahler component  ${\cal M}_{\rm 
  QK}$. 

Geometrically, this formulation involves going to an extension of
Hitchin's generalised geometry, called ``extended'' or ``exceptional
generalised'' geometry (EGG)~\cite{Chris,EGG}. 
In conventional generalised geometry the internal metric and $B$-field
degrees of freedom are ``geometrised'' by considering structures on
the generalised tangent space $TM\oplus T^*M$. In EGG, one further
extends the tangent space, so as to completely geometrise all the
degrees freedom including the RR fields $C^-$ and the
four-dimensional axion-dilaton $(\phi,\tB)$, as structures on
this larger ``exceptional'' generalised tangent space. 

This section is arranged as follows. In section~\ref{app:Tsub}
we give some basic $\Es7$ definitions and in~\ref{EGG} we briefly
discuss the structure of the EGG relevant to type IIA
compactifications to four dimensions.  This formalism leads to the
expectation that the moduli spaces $\MSK^+$ and $\MQK$ should be
cosets of the form $\Es7/H$. This is discussed in sections~\ref{hyper}
and \ref{vectors} 
for the hyper- and vector multiplets respectively. In~\ref{sconformal}
we briefly review  some properties of the superconformal compensator
formalism  which is related to the hyperk\"ahler cone construction
discussed in \ref{structures} and \ref{HKC}. Given the known
properties of homogeneous spaces, we argue in \ref{structures}  what
form the coset describing the hypermultiplet moduli spaces should
take. In section \ref{HKC} we show it explicitly by specifying the
embedding of the NS and RR degrees of freedom, and give an $\Es7$
invariant expression for the hyperk\"ahler potential $\chi$ 
following the explicit construction of ref.~\cite{KS-Wolf}, showing as
well its consistency  with the literature \cite{RVV}. In
section~\ref{vectors} we turn to the vector multiplet moduli space
${\cal M}_{\rm SK}^+$. In \ref{structuresvector} we argue what coset
it should correspond to,  and in \ref{SKE7} we give its explicit
construction. We  show  that there is indeed a natural special
K\"ahler metric, generalising the construction of~\cite{Hitchin}, and
that the corresponding K\"ahler potential is given by the
square-root of the $\Es7$ quartic invariant in complete analogy to the
Hitchin function in the $O(6,6)$ case. In section \ref{compat} we
discuss the hyper and vector-multiplet sectors and their compatibility.


\subsection{Basic $\Es7$ group theory}
\label{app:Tsub}

The group $\Es7$ can be defined in terms of its fundamental
56-dimensional representation. It is the subgroup of $\Symp(56,\bbR)$
which preserves, in addition to the symplectic structure $\cS$, a
particular symmetric quartic invariant $Q$.  

In order to make the connection to the generalised geometry formalism
it will be useful to study the decomposition under
 \beq
\Es7\supset\Tsub\ ,
\eeq
where $\Oo(6,6)$ corresponds to T-duality symmetry of generalised
geometry, while $\SL(2,\bbR)$ is the S-duality symmetry. The latter 
acts on the  axion-dilaton $S=\tilde{B}+\ii\ee^{-2\phifour}$ (where
$\tilde B$ is the six-form dual to $B_{\mu\nu}$,  and $\ee^{-2\phifour}$ 
is the four-dimensional dilaton six-form defined in
(\ref{eq:phifour})) by fractional linear 
transformations.\footnote{This should not be confused with the
  S-duality in type IIB mixing the dilaton with the axion coming from
  the RR sector.} 
The fundamental representation decomposes as 
\begin{equation} 
\label{56}
\begin{aligned}
   \rep{56} & = (\rep{2},\rep{12}) + (\rep{1},{\rep{32}})\ ,\\
   \lambda & = \left( \lambda^{i\Ta},\, \lambda^+ \right)\ ,
\end{aligned}
\end{equation}
where $i=1,2$ labels the $\SL(2,\bbR)$ doublet while 
$\Ta=1,\dots,12$ labels the fundamental representation of
$O(6,6)$. $\lambda^+ $ denotes 
a 32-dimensional positive-chirality $O(6,6)$ Weyl spinor. 

The adjoint representation $\rep{133}$ decomposes as 
\begin{equation} 
\label{133}
\begin{aligned}
   \rep{133} &= (\rep{3},\rep{1}) + (\rep{1},\rep{66})
      + (\rep{2},\rep{32}')\ , \\
   \mu &= \left( \mu^{i}{}_j, \,\mu^{\Ta}{}_{\Tb}, \, \mu^{i-} \right) \ .
\end{aligned}
\end{equation}
This choice of spinor chiralities $\lambda^+, \mu^{i-}$ is  precisely
the one relevant for type IIA; the corresponding expressions for
type IIB would require a swap of the chiralities.  The $O(6,6)$ vector
indices $\Ta$ can be raised and lowered using the $O(6,6)$ metric
$\etac_{\Ta\Tb}$, while the $\SL(2,\bbR)$ indices can be raised and
lowered using the $\SL(2,\bbR)$ invariant anti-symmetric tensor
$\epsilon$, so that for any given doublet $v^i$ we define
$v_i=\epsilon_{ij}v^j$ where $\epsilon_{12}=1$ and
$v^i=\epsilon^{ij}v_j$ with $\epsilon^{12}=-1$.     

The $\Es7$ symplectic and quartic invariants are given by 
\begin{equation}
\label{invs}
\begin{aligned}
   \cS(\lambda,\lambda') &=  \epsilon_{ij}\lambda^i\cdot\lambda'^j
      + \mukai{\lambda^+}{\lambda^{\prime+}} \ , \\
   Q(\lambda) &= 
      \tfrac{1}{48} \mukai{\lambda^+}{\Gamma_{\Ta\Tb}\lambda^+}
         \mukai{\lambda^+}{\Gamma^{\Ta\Tb}\lambda^+}
         \\ & \qquad \qquad
      - \tfrac{1}{2}\epsilon_{ij}\lambda^i_\Ta\lambda^j_\Tb
         \mukai{\lambda^+}{\Gamma^{\Ta\Tb}\lambda^+}
      + \tfrac{1}{2}\epsilon_{ij}\epsilon_{kl}
         \left(\lambda^i\cdot\lambda^k\right)
         \left(\lambda^j\cdot\lambda^l\right) \ ,
\end{aligned}
\end{equation}
where $X\cdot Y=\eta_{\Ta\Tb}X^\Ta Y^\Tb$ and $\Gamma^{\Ta\Tb}$ is the
antisymmetrised product of $O(6,6)$ gamma-matrices. The action of the
adjoint representation (with parameter $\mu$) which leaves these
invariant is given by  
\begin{equation} \label{adjfund}
\begin{aligned}
   \delta\lambda^{iA} &= \mu^i{}_j \lambda^{jA} + \mu^A{}_B\lambda^{iB}
      + \mukai{\mu^{i-}}{\Gamma^A\lambda^+} \ , \\
   \delta\lambda^+ &= \tfrac{1}{4}\mu_{AB}\Gamma^{AB}\lambda^+
      + \epsilon_{ij}\lambda^{iA}\Gamma_A \mu^{j-} \ . 
\end{aligned}
\end{equation}
The adjoint action on the $\rep{133}$ representation (with parameter
$\mu^\prime$) is given by $\delta\mu=[\mu',\mu]$ where 
\begin{equation} \label{adjac}
\begin{aligned}
   \delta \mu^{ i}{}_j 
      &= \mu^{\prime i}{}_k \mu^k{}_j - \mu^i{}_k\mu^{\prime k}{}_j
         + \epsilon_{jk}\left( \mukai{\mu^{\prime i-}}{\mu^{k-}}
         - \mukai{\mu^{i-}}{\mu^{\prime k-}} \right) \ , \\
 \delta  \mu^{ A}{}_B 
      &= \mu^{\prime A}{}_C \mu^C{}_B - \mu^A{}_C \mu^{\prime C}{}_B
         + \epsilon_{ij}\mukai{\mu^{\prime i-}}
            {\Gamma^A{}_B \mu^{j-}} \ , \\
   \delta \mu^{ i-} 
      &= \mu^{\prime i}{}_j \mu^{j-} - \mu^i{}_j \mu^{\prime j-}
         + \tfrac{1}{4}\mu'_{AB}\Gamma^{AB}\mu^{i-}
         - \tfrac{1}{4}\mu_{AB}\Gamma^{AB}\mu^{\prime i-} \ . 
\end{aligned}
\end{equation}
One can also define the invariant trace in the adjoint representation
\begin{equation}
\label{eq:adj-tr}
   \tr \mu^2 = \tfrac{1}{2}\mu^i{}_j\mu^j{}_i 
      + \tfrac{1}{4}\mu^A{}_B\mu^B{}_A 
      + \epsilon_{ij}\mukai{\mu^{i-}}{\mu^{j-}} \ . 
\end{equation}

Let us briefly mention a different decomposition of $\Es7$. The
maximal compact subgroup of $\Es7$ is $\hg$. In particular, in the
supersymmetry transformations, the two type II spinors really
transform in the fundamental representation under (the double cover)
$\SU(8)$. The fundamental and the adjoint representation
of $\Es7$ decompose under $\hg$ as
\begin{equation}
\label{eq:hg-decomp}
\begin{aligned}
 \rep{56} &= \rep{28} + \rep{\bar{28}} \ , \qquad \qquad  \rep{133} =\rep{63}+\rep{35}+\rep{\bar{35}} \ , \\
   \lambda &= ( \lambda^{\alpha\beta}, \bar{\lambda}_{\alpha\beta} ) \
, \qquad \qquad\
   \mu = ( \mu^\alpha{}_\beta, \mu^{\alpha\beta\gamma\delta},
            \smu_{\alpha\beta\gamma\delta} ) \ , 
 \end{aligned}
\end{equation}
where $\alpha=1,\dots,8$ denotes the fundamental of $\SU(8)$ and
where $\lambda^{\alpha\beta}$ and $\mu^{\alpha\beta\gamma\delta}$ are
totally antisymmetric, $\mu^\alpha{}_\alpha=0$ and $(*\bar
\mu)_{\alpha\beta\gamma\delta}=\mu_{\alpha\beta\gamma\delta}$ (indices
are raised and lowered with the $\SU(8)$ Hermitian metric, constructed
from the spinor conjugation matrix). 
The action of the adjoint representation on the fundamental 
representation is given by 

\begin{equation}
\label{spinorE7}
\begin{aligned}
   \delta \lambda^{\alpha\beta} &=
     \mu^\alpha{}_\gamma \lambda^{\gamma\beta} +
     \mu^\beta{}_\gamma \lambda^{\alpha\gamma} +
     \mu^{\alpha\beta\gamma\delta} \bar{\lambda}_{\gamma\delta} , \\
   \delta\bar{\lambda}_{\alpha\beta} &=
     - \mu^\gamma{}_\alpha \bar{\lambda}_{\gamma\beta} -
     \mu^\gamma{}_\beta \bar{\lambda}_{\alpha\gamma} +
     \bar{\mu}_{\alpha\beta\gamma\delta} \lambda^{\gamma\delta} ,
\end{aligned}
\end{equation}
Although we will not give it here, one can use six-dimensional gamma
matrices to give explicit relations between the $\hg$ and $\Tsub$
decompositions.

\subsection{EGG for type IIA with $\Es7$}
\label{EGG}

To define exceptional generalised geometry for type IIA compactified
to four dimensions, one starts with an extended generalised tangent
space (EGT) of the form\footnote{This structure was first discussed 
  in~\cite{Chris} and, in an M-theory context, in~\cite{EGG}. For a
  more complete description of the geometry in this particular case 
  see~\cite{aaron}.}  
\begin{equation}
\label{eq:EGT}
   E  = TM \oplus T^*M 
            \oplus \Lambda^5T^*M 
            \oplus \left(T^*M\otimes\Lambda^6T^*M\right) 
            \oplus \Lambda^\textrm{even}T^*M \ .
\end{equation}
The first two terms correspond to the generalised tangent
bundle of conventional generalised geometry and are loosely associated
to the momentum and winding of string states. The next two terms can be
thought of corresponding to NS five-brane and Kaluza--Klein monopole
charges. The final term is isomorphic to $S^+$, the positive helicity
$\Spin(6,6)$ spinor bundle, and is associated to D-brane charges. 
The EGT space is  56-dimensional and, just as there was a natural
$O(d,d)$-invariant metric on $TM\oplus T^*M$, there is a natural
symplectic form $\cS$ and symmetric quartic invariant $Q$ on
$E$.  The group that preserves both $\cS$ and $Q$
is $\Es7$. Thus the analogue of the $O(d,d)$ action is a natural $\Es7$ action on
$E$. Essentially~\eqref{eq:EGT} corresponds to a decomposition 
of the $\rep{56}$ fundamental representation of $\Es7$ under a
particular $\GL(6,\bbR)\subset\Es7$ which is identified with
diffeomorphisms of $M$. The embedding of $\GL(6,\bbR)$ in $\Es7$ is described explicitly
in Appendix \ref{app:EGG}. More precisely, as discussed there,
$E$ corresponds to the decomposition of the fundamental representation weighted
by $(\Lambda^6T^*M)^{1/2}$.

One can similarly decompose the adjoint $\rep{133}$ representation of $\Es7$ under
this $\GL(6,\bbR)$ subgroup and finds (see~\eqref{eq:adj6}) 
\begin{equation}
\begin{aligned}
   A_0 &= \left(TM\otimes T^*M\right)
            \oplus \Lambda^2TM \oplus \Lambda^2T^*M
            \\ & \qquad \qquad 
            \oplus \bbR \oplus \Lambda^6T^*M \oplus \Lambda^6TM
            \oplus \Lambda^\textrm{odd}T^*M
            \oplus \Lambda^\textrm{odd}TM \ .
\end{aligned}
\end{equation}
The first term corresponds to the $\GL(6,\bbR)$ action. However, there
is a second important subgroup one can obtain from taking only the
$p$-form elements of $A_0$
\begin{equation}
\label{eq:forms}
   B+\tB+C^- \in \Lambda^2T^*M \oplus\Lambda^6T^*M
       \oplus\Lambda^\textrm{odd}T^*M  \ ,
\end{equation}
giving a nilpotent sub-algebra~\eqref{form-alg}. These are the EGG
analogues of the ``$B$-shift'' symmetries of generalised geometry and
are in one-to-one correspondence with the form-fields of the IIA
supergravity. In particular, $B$ is the internal $B$-field, $C^-$ the
RR-form potentials and $\tB$ is an internal six-form corresponding to
the ten-dimensional dual of $B_{\mu\nu}$. To identify $B$, $\tilde B$
and $C^-$ in the $\Tsub$ decomposition of the adjoint representation
(\ref{133}), we note, as explained in detail in Appendix
\ref{app:EGG}, that the embedding of $\GL(6,\bbR)\subset \Tsub
\subset\Es7$ breaks the $SL(2,\bbR)$ symmetry, picking out a
$\SL(2,\bbR)$ vector $v^i$. Using this vector, we identify $B$,
$\tilde B$ and $C^-$ as the following elements 
in the $\rep{133}$ 
\begin{equation}
\label{potentials}
\begin{aligned}
   \mu^i{}_j &= \tB_{1\dots 6} v^i v_j  \ , &&&
      \tB &\in \Lambda^6T^*M \ , \\
   \mu^A{}_B &= \begin{pmatrix} 0 & 0 \\ B & 0 \end{pmatrix} \ , &&&
      B &\in \Lambda^2T^*M \ , \\
   \mu^{i-} &= v^i C^- \ , &&&
      C^- &\in \Lambda^{\textrm{odd}}T^*M \ . 
\end{aligned}
\end{equation}

Geometrically, these form-field potentials together with the metric
and dilaton encode an $\hg$ structure on $E$~\cite{EGG}, i.e.~they
parameterise the coset $\Es7/(\hg)$. Formally this structure is an 
element $I\in\Es7$ that, like a complex structure, satisfies
$I^2=-\id$. This then defines a (exceptional generalised) metric (EGM)
on $E$ given by $G(\lambda,\lambda)=\mathcal{S}(\lambda,I\lambda)$
where $\lambda \in E$. This is the analogue of the generalised
metric on $TM\oplus T^*M$. One can show that a generic EGM can
be written as\footnote{The choice of sign for $B$ is conventional, to
  match the usual generalised geometry $B$-shift.} 
\begin{equation}
\label{EGM}
   G(\lambda,\lambda) 
      = G_0(\ee^{C^-}\ee^{\tB}\ee^{-B}\lambda,
          \ee^{C^-}\ee^{\tB}\ee^{-B}\lambda) \ , 
\end{equation}
where $G_0$ is a specific EGM built from $g$ and the dilaton $\phi$,
the form of which will not be important, and
$\ee^{C^-} \ee^{\tB}\ee^{-B}$ are the exponentiated actions of the adjoint elements given
in~\eqref{eq:forms}. Hence $C^-$, $\tB$, $B$ $g$ and $\phi$ encode a 
generic EGM, or equivalently a point in the coset $\Es7/(\Tsub)$. 

If the form field strengths are nontrivial, the potentials  $B$, $\tB$
and $C^-$ can only be defined locally. The EGM is then really a metric
on a twisted version of~\eqref{eq:EGT}, where we introduce on each
patch $U_{(\alpha)}$  
\begin{equation}
   \lambda_{(\alpha)} = \ee^{C^-_{(\alpha)}}
      \ee^{\tB_{(\alpha)}}\ee^{-B_{(\alpha)}}\lambda,
\end{equation}
such that on $U_{(\alpha)}\cap U_{(\beta)}$ we have a patching by
gauge transformations 
\begin{equation}
   \lambda_{(\alpha)} = \ee^{\dd\Lambda^+_{(\alpha\beta)}}
      \ee^{\dd\tL_{(\alpha\beta)}}
      \ee^{-\dd\Lambda_{(\alpha\beta)}}\lambda_{(\beta)} \ , 
\end{equation}
which implies
\begin{equation}
\label{gauge-transf}
\begin{aligned}
   B_{(\alpha)} &= B_{(\beta)} + \dd\Lambda_{(\alpha\beta)} \  , \\
    C^-_{(\alpha)} &= C^-_{(\beta)} + \dd\Lambda^+_{(\alpha\beta)} 
       + \ee^{-\dd\Lambda_{(\alpha\beta)}}C^-_{(\beta)} \ , \\
   \tB_{(\alpha)} &= 
       \tB_{(\beta)} + \dd\tL_{(\alpha\beta)} 
       + \mukai{\dd\Lambda^+_{(\alpha\beta)}}
          {\ee^{-\dd\Lambda_{(\alpha\beta)}}C^-_{(\beta)}} \  . 
  \end{aligned}
\end{equation}
These correspond precisely to the gauge transformations of the
relevant supergravity potentials. Comparing with~\eqref{GCdef}, we
see, in particular, that the field strengths $H=\dd B$ and
$F=\sqrt{2}\ee^B \dd C^-$ are gauge invariant. The transformation of
$\tB$ similarly matches the form given in~\cite{Bdual}, specialised to
six dimensions. 

Having summarised the key components of the EGG, let us now turn to
the issue of how this structure can be used to describe the
hypermultiplet and vector multiplet sectors. 


\subsection{Hypermultiplet sector}
\label{hyper}

$N=2$ supergravity constrains the scalar degrees of
freedom in the hypermultiplets to span a  quaternionic-K\"ahler manifold
${\cal M}_{\rm QK}$. Over any such manifold one can construct a
hyperk\"ahler cone  ${\cal M}_{\rm HKC}$ which has one additional
quaternionic dimension~\cite{Swann,HKC}. In the following section we
briefly review the appearance of K\"ahler cones in superconformal
supergravity.  The metric on the cone is characterised by a single
function $\chi$ known as the hyperk\"ahler potential. In
section~\ref{HKC} we then identify how the NS and RR degrees of
freedom can be embedded into an $\Es7$ EGG structure.
We show that they parameterise a coset known as a ``Wolf
space''~\cite{wolf,alex}, which admits a standard construction of a
hyperk\"ahler cone~\cite{KS-Wolf}, with an $\Es7$ invariant
expression for the hyperk\"ahler potential. 

\subsubsection{Hyperk\"ahler cones and superconformal supergravity}
\label{sconformal}

Superconformal supergravity has as  the space-time symmetry
group the superconformal group instead of the super-Poincare
group. Using a compensator formalism one can  
construct superconformally invariant actions and then obtain Poincare
supergravity 
as an appropriately  gauge fixed version.
We cannot review the entire subject here but
let us recall the properties relevant for our 
subsequent discussion following refs.~\cite{HKC,VPnotes}.

In the case of $N=2$ one adds a compensating vector multiplet and a
compensating hypermultiplet to the spectrum and couples all multiplets
to the Weyl supermultiplet which contains the gravitational degrees of
freedom.  One of the resulting features is that the 
$N=2$ R-symmetry $\SU(2)_{\rm{R}}\times U(1)_{\rm{R}}$  together with the dilation symmetry are gauged. 
Furthermore, 
the dimension of the scalar manifolds are enlarged
by one `unit' and its geometry is altered. For the hypermultiplets this 
precisely corresponds to the hyperk\"ahler cone construction where
the four additional scalar fields of the compensator can be viewed as forming
a cone (with one radial direction and an $S^3$) over the quaternionic-K\"ahler base  ${\cal M}_{\rm QK}$.
The geometry of this cone is no longer quaternionic-K\"ahler but instead
hyperk\"ahler in that the three local almost complex structures of ${\cal
  M}_{\rm QK}$ lift to globally defined integrable structures on the
cone.  Conversely, a quaternionic-K\"ahler manifold can be viewed as a 
quotient 
\beq\label{HKcone}
 {\cal M}_{\rm QK}\ =\ \frac{{\cal M}_{\rm HKC}}{\SU(2)_{\rm R}\times \bbR^+} \
,
\eeq
where $\bbR^+$ corresponds to the dilatations and 
the $\SU(2)_{\rm R}$ rotates the three almost complex structures of 
 ${\cal M}_{\rm QK}$.

${\cal M}_{\rm HKC}$ can be characterised by a hyperk\"ahler potential
$\chi$ which is simultaneously  a K\"ahler potential for all three
complex structures. A generic expression for $\chi$ in terms of the three complex structures was given in
\cite{Swann,HKC}, while for the specific case of hyperk\"ahler cones
which arise from a special geometry via the c-map, $\chi$ was
determined in refs.~\cite{RVV,pioline}. In this case a particularly
simple expression results in a gauge where the $\SU(2)_{\rm R}$ is partially
fixed to a $U(1)_{\rm R}$ subgroup \cite{RVV} and one finds\footnote{The expression for $\chi$ in an arbitrary gauge is given in \cite{pioline}.}
\beq\label{chiRVV}
\chi\ =\ G^{-1}_0\, \ee^{-K_{\rm SK}} \ ,
\eeq
where $K_{\rm SK}$ is the K\"ahler potential of special K\"ahler
subspace  ${\cal M}_{\rm SK}$
and $G_0$ contains the dilaton $\phi$ together with the compensator corresponding to the cone
direction. More precisely,\footnote{Note 
that compared to ref.~\cite{RVV} we have a different convention of the
dilaton. The dilaton used in that paper is obtained from the dilaton
here by the replacement $2\phi \to -\phi$.} 
here $K_\textrm{SK}=K^-$ and 
\beq\label{Gzero}
G^{-1}_0 = \tfrac12 \, \ee^{\frac12 K^- - \phi} \  ,
\eeq
where in this parameterisation the compensator for the dilatations and
the $\U(1)_{\rm R}$ are identified with the degrees of freedom in the
pure spinor $\Phi^-$ which correspond to the complex rescaling $\Phi^- \to
c \Phi^-$.\footnote{Note that this expression for the compensator also 
appears in the $N=1$ orientifolded version analysed in ref.~\cite{GL}.}
Inserting \eqref{Kahlerpot} one finds 
\beq\label{chiVandoren}
\chi\ =\ \tfrac12  \ee^{-\phi} \, \sqrt{\ii \mukai{\Phi^-}{\bar\Phi^-}}\ .
\eeq

On the HKC also the expression for the Killing prepotentials $\cP_a$
change.  In ref.~\cite{HKC} it was found 
\begin{equation} 
\begin{aligned} \label{PHKC}
   \cP_\pm^{\rm HCK} \ =\ \chi\, \ee^{\pm\ii\alpha} \, \cP_\pm \ ,
   \qquad
   \cP_3^{\rm HCK}\ =\  \chi \,\cP_3 \ ,
   \end{aligned}
\end{equation}
where $\ee^{\ii\alpha}$ parameterises the angle variable of the
$\U(1)_{\rm R}$, and $\cP_a$ are the Killing prepotentials on the
quaternionic space. In the notation used above $\ee^{\ii\alpha}$ is
the phase of the scale parameter $c$.


\subsubsection{Expected coset for hypermultiplet sector}
\label{structures}

As argued before, we  expect the moduli space $\MQK$ to be a coset of the form $\Es7/H$,
corresponding to defining a particular structure $H$ on the
exceptional generalised tangent space $E$. 
Given the coset structure of $\MSK^-$ displayed in \eqref{MSK},
and the fact that this moduli space is related to $ {\cal M}_{\rm QK}$ by the c-map ${\cal M}_{\rm
  SK}^-\to {\cal M}_{\rm QK}$,
we can
actually make a simple conjecture for the form of
$\mathcal{M}_{\text{QK}}$.  For
the case of special K\"ahler coset spaces the corresponding
quaternionic spaces are known \cite{CFG,cecotti,dWVP}. For our
particular case, we learn that the c-map relates 
\beq\label{cmap}
\textrm{c-map}: \qquad {\cal M}_{\rm SK}^- 
   = \frac{\Oo(6,6)}{U(3,3)} \
   \to\ {\cal M}_{\rm QK} = \frac{\Es7}{\SO^*(12) \times SU(2)}\ . 
\eeq
The map is usually given for the compact real versions of these groups
so we have actually generalised slightly to consider particular
non-compact forms\footnote{The non-compact group
  $\SO^*(2n)$ is a real form of the complex group $\SO(2n,\bbC)$ where
  the elements of the corresponding Lie algebra are complex matrices
  of the form $\left(\begin{smallmatrix}A & B \\ -B^* &
        A^*\end{smallmatrix}\right)$ with $A=-A^T$ and
  $B=B^\dag$. For more details see for instance~\cite{BB}.} giving a
pseudo-Riemannian metric of signature $(40,24)$. The compact version
of ${\cal M}_{\rm QK}$ is one of the well-known Wolf spaces~\cite{wolf,alex}
and has $\textrm{dim}({\cal M}_{\rm QK})= 64$. We see that as
anticipated the U-duality group $\Es7$ determines the geometry of the
quaternionic-K\"ahler space, and corresponds to the space of
$\SO^*(12)\times\SU(2)$ structures on $E$. Furthermore, the dimension
64 precisely matches the expected supergravity hypermultiplet degrees
of freedom: 30 in $\Phi^+$ (since it is defined modulo complex
rescalings), 32 in $C^+$ and two more in $\phi$ and $\tB$.   
The hyperk\"ahler cone corresponding to the Wolf space given in
\eqref{cmap} is the space
\beq\label{Wcone}
   {\cal M}_{\rm HKC}\ =\ \frac{\Es7}{\SO^*(12)}\times \bbR^+ \ ,
\eeq
with $\textrm{dim}({\cal M}_{\rm HCK})= 68$. This space has been
studied very explicitly in the mathematical literature and in
particular a hyperk\"ahler potential $\chi$ has been
given~\cite{KS-Wolf}.  Let us now use this construction
to verify our expectations.


\subsubsection{Hyperk\"ahler cone construction \`a la Swann}
\label{HKC}


In this section we show explicitly how $\Phi^-$, $C^-$ together with
the dilaton/axion pair $(\phi,\tilde B)$ parameterise the 64-dimensional 
Wolf space $\MQK = \Es7/(\SO^*(12)\times\SU(2))$, give the
construction of the corresponding  hyperk\"ahler cone metric on $\MHK$
following~\cite{KS-Wolf} and derive the form of the hyperk\"ahler 
potential $\chi$. 

The analysis of compact symmetric spaces $G/H$ with quaternionic
geometry is due to Wolf~\cite{wolf} and Alekseevskii~\cite{alex} who
showed there is one such space for each compact simple Lie
group. Swann~\cite{Swann} subsequently identified the corresponding
hyperk\"ahler cone structures, viewing $\MHK$ as an orbit in the adjoint
representation under the complexified $G$. Koback and Swann then gave
an explicit expression of the hyperk\"ahler cone \cite{KS-Wolf}. 
Here, we will follow these constructions to give an explicit form of
the quaternionic geometry on $\Es7/(\SO^*(12)\times\SU(2))$ in terms
of the supergravity degrees of freedom.  

The hyperk\"ahler cone $\MHK$ can be viewed as an orbit in the
$\rep{133}$ adjoint representation in two ways. In the complexified
version one starts with an element $\JJ_+\in\mathfrak{e}_7^\bbC$
corresponding to a highest weight root in the Lie algebra. The space
$\MHK$ is then the orbit of $\JJ_+$ under $E_7^\bbC$. In this picture
$\JJ_+$ is stabilised under 99 elements of $E_7^\bbC$ so that $\MHK$
is a $133-99=34$ complex-dimensional space. Given a real structure,
which for us means the non-compact real form $\Es7$, one can identify
the complex conjugates of elements of $\rep{133}$. This defines
$\JJ_-=\bar{\JJ}_+$ and hence, for each $\JJ_+$ in the orbit, a
particular $\su(2)$ subalgebra in the real algebra $\es7$  generated
by $\JJ_\pm$ and the corresponding $\JJ_3\sim [\JJ_+,\JJ_-]$. In this
second picture $\MHK$ is the orbit of this $\su(2)$ algebra, under the real
group $\Es7$, together with an overall  scaling, where the triplet
$\JJ_a$ is stabilised by a 66-dimensional $\SO^*(12)$ subgroup of
$\Es7$. The overall scaling of $\JJ_a$ represents the radial direction
of the hyperk\"ahler cone, while the $\SU(2)$ action on the cone is
realized by the action of the $\su(2)$ algebra on itself, rotating the
triplet $\JJ_a$. 

As discussed in section~\ref{app:Tsub}, under $\Tsub$ the adjoint
representation of $\Es7$ decomposes as $   \mu = \left( \mu^{i}{}_j,
\,\mu^{\Ta}{}_{\Tb}, \, \mu^{i-} \right)$ corresponding to  
$\rep{133} = (\rep{3},\rep{1}) + (\rep{1},\rep{66})
+ (\rep{2},\rep{32}')$ (see~\eqref{133}).
Given an $\SU(3,3)$ structure $\Phi_0^-$ as defined in (\ref{purespinors}), we can then identify a triplet
of elements, where $\JJ_\pm^{(0)}=\JJ_1^{(0)}\pm\ii \JJ_2^{(0)}$,   
\begin{equation}
\label{JE7}
\begin{aligned}
   \JJ_+^{(0)} &= \left( 0,0,  u^i\Phi_0^- \right) \ , \\ 
   \JJ_-^{(0)} &= \left( 0,0,  \bar{u}^i\bar{\Phi}_0^- \right) \ , \\ 
   \JJ_3^{(0)} &= \tfrac{1}{4}\ii\kappa^{-1}\mukai{\Phi^-}{\bar{\Phi}^-} 
        \left( (u^i\bar{u}_j+\bar{u}^iu_j) , (\ii u\bar{u})
           \cJ_0^{-\Ta}{}_\Tb, 0\right) \ .
\end{aligned}
\end{equation}
We have also used the
fact that $\mukai{\Phi^-_0}{\bar{\Phi}^-_0}=\mukai{\Phi^-}{\bar{\Phi}^-}$ and
defined 
\begin{equation}
\label{kdef}
   \kappa = \sqrt{ \tfrac{1}{2}\ii\mukai{\Phi^-}{\bar{\Phi}^-}
        (-\ii u\bar{u}) } \ .
\end{equation}
Here $u^i$ is a complex vector transforming as a doublet under
$\SL(2,\bbR)$, $\cJ_0^-$ is the generalised complex
structure~\eqref{Jgen} defined by $\Phi_0^-$ and we abbreviate
$uv=\epsilon_{ij}u^jv^i=u_iv^i=-u^iv_i$. The triplet
$(\JJ_1^{(0)},\JJ_2^{(0)},\JJ_3^{(0)})$ then satisfies the (real)
$\su(2)$ algebra   
\begin{equation}
\label{SU2}
   \big[ \JJ_a^{(0)}, \JJ_b^{(0)} \big] 
      = 2\kappa\,\epsilon_{abc} \JJ_c^{(0)}\ .
\end{equation}
We have included an overall scaling $\kappa$ in the $\su(2)$ algebra
since, as mentioned above, this corresponds to the radial direction on
the hyperk\"ahler cone. 

We would now like to see the action of $\Es7$ on the triplet
$\JJ_a^{(0)}$ to find the dimension of the corresponding orbit. In
particular we should find that the triplet is stabilised by a
66-dimensional subgroup of $\Es7$. We first note that, by definition,
$\Phi_0^-$ and hence $\cJ_0^-$ are invariant under $\SU(3,3)\in
O(6,6)$ which correspond to 35 stabilising elements. 
There are no elements of $\sll(2,\bbR)$ which leave $u^i$
invariant, though the element
\begin{equation}
\label{R-element}
   \left( u^i\bar{u}_j+\bar{u}^iu_j,
         -\tfrac{1}{3}(\ii u\bar{u})\cJ_0^{-\Ta}{}_\Tb,0 \right) 
\end{equation}
in $\sll(2,\bbR)\times \so(6,6)$ does commute with all three
$\JJ^{(0)}_a$, and also with the $\SU(3,3)$ action. Finally we have
the action of elements of the form $(0,0,\mu^{i-})$. Without loss of
generality we can write $\mu^{i-}=u^i\mu^-+\bar{u}^i\bar{\mu}^-$, 
and then find using (\ref{adjac}) that, to
be a stabiliser, $\mu^-$ is required to satisfy
\begin{equation}
\label{alpha-cond}
   \mukai{\Phi_0^-}{\mu^-} = \mukai{\Phi_0^-}{\bar{\mu}^-} = 0 
   \ , \qquad  
   \tfrac{1}{4}\cJ^-_{0\,\Ta\Tb}\Gamma^{\Ta\Tb}\mu^- = \ii\mu^- \ . 
\end{equation}
Under the $\SU(3,3)$ group defined by $\Phi_0^-$, the $\rep{32}'$
spinor representation decomposes as
$\rep{1}+\rep{1}+\rep{15}+\bar{\rep{15}}$. The
conditions~\eqref{alpha-cond} imply that $\mu^-$ is in the
$\rep{15}$ representation, and hence we see there are a further 30
real elements in $\es7$ which stabilise the $\JJ_a^{(0)}$. Thus
together with the $\su(3,3)$ algebra and the element~\eqref{R-element}
we see that the stabiliser group is 66 dimensional. It is relatively
straightforward to show that this group has signature $(30,36)$ and hence
corresponds to $\SO^*(12)$. 

We now address how to generate a generic element in the orbit from
the specific  $\JJ_a^{(0)}$ discussed so far. We first note
that the $O(6,6)\subset \Es7$ transformations of $\Phi_0^-$ by $B$ as
in (\ref{purespinors})  generate the full $O(6,6)$ orbit.  These
transformations are embedded in $\Es7$ as in (\ref{potentials}). On
the other hand, $\SL(2,\bbR)$ elements simply rotate $u^i$, which was
already assumed to be general. Apart from the $\SU(2)$ rotations
among the $\JJ_a^{(0)}$, the only additional motion in the orbit
comes from elements of the form $(0,0,\mu^{i-})$. We expect that
these should correspond to the RR scalars $C^-$. To see this
explicitly, we first recall that 30 of these  leave $\JJ_a^{(0)}$
invariant, while anything of the form
$\mu^{i-}=A\re(u^i\Phi^-)+B\im(u^i\Phi^-)$ simply generates part of
the $\SU(2)$ rotations among the triplet $\JJ^{(0)}_a$. The remaining
32 degrees of freedom can be generated by acting with the RR 
potential $C^-$ embedded in $\Es7$ as in~\eqref{potentials}, since it
is easy to show that none of these elements leave $\JJ_a^{(0)}$
invariant. Hence the generic triplet in the orbit can be written as  
\begin{equation}
\label{Jgeneric}
   \JJ_{a} = \ee^{C^-} \ee^{\tB} \ee^{-B} \JJ_a^{(0)\prime}\ ,
\end{equation}
where $\JJ_a^{(0)\prime}$ are the $\SU(2)$ rotation of
$\JJ_a^{(0)}$, that we can parameterise as 
\begin{equation}
\label{J-angle}
\begin{aligned}
   \JJ^{(0)\prime}_3 &=  \tfrac12 \sin\theta \,\ee^{\ii\alpha} \, \JJ^{(0)}_+ 
                 +  \tfrac12 \sin\theta \,\ee^{-\ii\alpha} \, \JJ^{(0)}_- 
                 - \cos\theta \, \JJ^{(0)}_3 \ , \\
   \JJ^{(0)\prime}_+ &= \tfrac12 (1-\cos\theta)\ee^{\ii(\psi+\alpha)}\JJ^{(0)}_+ 
                 -  \tfrac12 (1+\cos\theta ) \ee^{\ii(\psi-\alpha)}\JJ^{(0)}_- 
                 -\ee^{\ii\psi} \sin\theta \JJ^{(0)}_3 \ .
\end{aligned}
\end{equation}
Note that the angle $\alpha$ corresponds to an $U(1) \subset \SU(2)$
phase rotation on $\JJ_+^{(0)}$, which can be absorbed in $\Phi_0^-$.  
Similarly the $\ee^{\tB}$ action is in $\SL(2,\bbR)$ and 
so is strictly speaking
unnecessary since it can be absorbed in a redefinition of $u^i$.
However it is useful to include it to see the structure of how the
supergravity potentials appear. 
To see that the orbit is 68-dimensional we note that $\Phi^-$ and
$C^-$ each contributes $32$ degrees of freedom. In the original ansatz
we can always rescale $\Phi^-\to c\Phi^-$ and $u^i\to c^{-1}u^i$ for
$c\in\bbC-\{0\}$ so there are really only two new real degrees of freedom
in $u^i$. In addition there are two degrees of freedom in $\theta$ and
$\psi$ giving a total of 68. 

Having given an explicit parameterisation of the coset space, we can
now consider the hyperk\"ahler structure and hyperk\"ahler potential
following ref.~\cite{KS-Wolf}. The result for the latter is very
simple: at a generic point on $\MHK$ it is given by  
\beq\label{chiKS}
\chi\ = \sqrt{-\tfrac{1}{8}\tr(\JJ_+\JJ_-)} \ ,
\eeq
where the trace is defined in~\eqref{eq:adj-tr}. Inserting \eqref{JE7},
\eqref{Jgeneric} and \eqref{J-angle} we find that 
\beq\label{chiWolf}  
   \chi = \sqrt{\tfrac18 (-\ii u\bar{u})\ii\mukai{\Phi^-}{\bar{\Phi}^-}} 
      =\sqrt{\tfrac18 (-\ii u\bar{u})H(\re \Phi^-)}\ ,
\eeq
where in the second equation we have used \eqref{Kahlerpot} and
\eqref{KrigidH}. Note that in fact $\chi=\tfrac12 \kappa$ where $\kappa$
was the normalization of the $\su(2)$ algebra~\eqref{SU2} and so is
manifestly an $\SU(2)$ invariant. Furthermore, it is independent of
$C^-$ since it is an $\Es7$-invariant function of $\JJ_a$, and
$e^{C^-}$ is an $\Es7$ transformation. Comparing
with~\eqref{chiVandoren} we see that the $\chi$ agree if we
identify\footnote{In a generic $\SU(2)$ gauge, i.e.\ using $\chi$ as 
  found in \cite{pioline}, we would get $ -\ii u \bar u\ = 2\ee^{-2\phi}
  \sin^2\theta$. This would not spoil the consistency verified in
  section  \ref{E7P} below, so for convenience we use the gauge fixed
  expression  \eqref{chiVandoren}. We thank B. Pioline for discussions
  on this point.} 
\beq\label{ident}
   -\ii u\bar{u} = 2\ee^{-2\phi} \ .
\eeq
In section \ref{E7P} we compute the Killing prepotentials which will
allow us to determine the dilaton dependence of $u$, providing an
independent confirmation of (\ref{ident}). 

In addition to the hyperk\"ahler potential there should be a triplet 
of complex structures $(I_1,I_2,I_3)$ acting on vectors in the tangent
space of the cone. Recall that a general point on the cone is defined
by the triplet $\JJ_a$, while a generic vector can be viewed as small
deformation $\delta\JJ_a$ along the cone. A general deformation around
the orbit is generated by the action of some $\mu\in\rep{133}$ on
$\JJ_a$. To fill out the full cone we also need to consider rescalings
of $\JJ_a$. Thus a vector in the tangent space of the cone at the
point $\JJ_a$ is a triplet that can be written as 
\begin{equation}
   \xi_a = [ \mu, \JJ_a ] + \mu_0 \JJ_a . 
\end{equation}
for some $\mu\in\rep{133}$ and $\mu_0\in\bbR^+$. Since $\JJ_a$ satisfy
the $\su(2)$ algebra~\eqref{SU2}, one only needs to specify two
elements (say $\JJ_1$ and $\JJ_2$ or equivalently $\JJ_+$) to
determine the triplet. Similarly, the vector in the tangent space
is completely determined by giving only two of the three $\xi_a$. The
three complex structures are then most easily defined by picking out 
different pairs of $\xi_a$ to specify the vector. In particular one
defines the structure $I_3$ by taking the vector defined by the pair
$(\xi_1,\xi_2)$ with the simple action 
\begin{equation}
   I_3 \left(\xi_1 + \ii\xi_2 \right) 
      = \ii\left(\xi_1 + \ii\xi_2 \right) \ ,
\end{equation}
with the corresponding cyclic relations defining $I_1$ and $I_2$. 


\subsection{Vector multiplets}
\label{vectors}

We now turn to the vector multiplet moduli space ${\cal  M}_{\rm
  SK}^+$. The superconformal supergravity formalism requires an
additional vector multiplet whose scalar degrees of freedom are the
conformal compensator corresponding to the overall scale of
$\Phi^+$. This adds a $U(1) \times \mathbb R^+$ factor to the moduli
space, turning the local special K\"ahler geometry $\MSK^+$ into a
rigid one $\Mrigid^+$. Both of them are expected to be cosets of the
form $\Es7/H$, up to an $\bbR^+$ factor. In the following section we
anticipate the form of the coset. In section \ref{SKE7} we identify
the embedding of the NS degrees of freedom into an orbit which spans the
expected coset, and show that the K\"ahler potential is given by the
square-root of the $\Es7$ quartic invariant in complete analogy to the
Hitchin function in the $O(6,6)$ case. 


\subsubsection{Expected cosets for vector multiplet moduli space}
\label{structuresvector}

It is well
known~\cite{dWVP} which coset manifolds have a local special K\"ahler
geometry and there is only one candidate based on $\Es7$ 
\begin{equation}
\label{vec-local}
   \MSK^+ = \frac{\Es7}{\Ex6\times U(1)} \ .
\end{equation}
(Again we are actually using a particular non-compact and
non-Riemannian version with signature $(30,24)$.) There is also the
corresponding rigid special K\"ahler space 
\begin{equation}
\label{vec-cone}
   \Mrigid^+ = \frac{\Es7}{\Ex6}\times\bbR^+ \ , 
\end{equation}
such that $\MSK^+=\Mrigid/\bbC^*$.

\subsubsection{Explicit construction}
\label{SKE7} 

We would like to see explicitly how $\Phi^+$ can be used to
parameterise the special K\"ahler coset spaces \eqref{vec-local} and
\eqref{vec-cone} 
and how the metric on each is defined in
terms of $\Es7$ objects. 

The space~$\Mrigid^+$ is actually what is known as a
``prehomogeneous'' vector space~\cite{SK}, that is, it is an open orbit of
$\Es7$ in the 56-dimensional representation. This is in complete
analogy to the $\Phi^\pm$ moduli spaces, which were open orbits in the
spinor representations $\rep{32}^\pm$ under $O(6,6)$. However, it is
in contrast to the hypermultiplet space $\MHK$ discussed above. For
us, the main point is that we should be able to realise the space as
the orbit of some embedding of $\Phi^+$ in the $\rep{56}$
representation. As discussed in section~\ref{app:Tsub}, under $\Tsub$,
the fundamental representation decomposes as $   \lambda = \left(
\lambda^{i\Ta},\, \lambda^+ \right)$  corresponding to 
$\rep{56}  = (\rep{2},\rep{12}) + (\rep{1},{\rep{32}})$.
 We would like to have a real orbit, so it is natural to start with an
embedding 
\begin{equation}
\label{lambz}
   \lambda^{(0)} = \left(0, \rePhi_0^+\right) \ ,
\end{equation}
where $\rePhi_0^+=2\re\Phi_0^+$. (The factor of two is chosen to 
match~\eqref{KrigidH}).

To check that this is a reasonable choice, we first note that it
implies that $\Phi^+$ is a singlet under the S-duality group  
$\SL(2,\bbR)\subset\Es7$. This is exactly what we would expect. In
section \ref{O66} we recalled that the dilaton is part of a
hypermultiplet, and therefore it should not couple to $\Phi^+$,
implying the latter is a singlet under S-duality. Alternatively, in type IIA
the $\U(1)_{\rm R}$ acts on the $\Spin(6)$ spinors by the phase
rotation $\eta^I_+\to\ee^{\ii\alpha/2}\eta^I_+$. This follows from
(\ref{decompepsilon}) together with the fact that the four-dimensional
supersymmetry parameters $\varepsilon^{1}_+$, $\varepsilon^2_+$
transform with opposite phases under the $\U(1)_{\rm R}$, while the
ten-dimensional $\epsilon^{1}$, $\epsilon^{2}$ are
invariant.\footnote{The same argument implies that in type IIB the
  spinors $\eta^I_+$ rotate with opposite phases.} Using
(\ref{purespinors}) we see that $\Phi^+$ is a singlet under
the $U(1)_\textrm{R}$ in type IIA while $\Phi^-$
rotates with a phase. This phase rotation is generated by
the $\SU(2)_R$ generator $\JJ_3$ defined in section~\ref{HKC}. Since
\eqref{JE7} shows the embedding of $\SU(2)_{\rm R}$ 
into $\Es7$, we conclude that $\Phi^+$ has to be  singlet under
the S-duality $\SLR$. Thus we see that  embedding $\Phi^+$ into the
$\rep{56}$ of $\Es7$ is also consistent with the action of the $N=2$
R-symmetry. 

In order to fill out the full orbit, we must act on $\lambda^{(0)}$
with $\Es7$ so 
\begin{equation}
   \lambda = g\cdot\lambda^{(0)} \ , \qquad g\in\Es7 \ .
\end{equation}
We can see that the dimension of
the orbit is indeed 56 by looking at the stabiliser of
$\lambda^{(0)}$. Using the decomposition~\eqref{133} we see from 
(\ref{adjfund}) that $\delta \lambda^{(0)}=0$ holds for the following
78 elements of $e_{7(7)}$: three from the $\SL(2,\bbR)$ elements
$\mu^i{}_j$, which do not act on $\lambda_{(0)}$; 35 from
$\mu^\Ta{}_\Tb$ since by construction $\re\Phi_0^+$ is stabilised by
$\SU(3,3)\subset O(6,6)$; and 40 from $\mu^{i-}$, since they must satisfy
$\mukai{\mu^{i-}}{\Gamma^\Ta\re\Phi_0^+}=0$, giving 24 conditions for 64
parameters $\mu^{i-}$. Put another way, decomposing under
$\SL(2,\bbR)\times\SU(3,3)\subset\Es7$, the adjoint action of the
stabiliser group transforms as $(\rep{3},\rep{1}) + (\rep{1},\rep{35})
+ (\rep{2},\rep{20})$, which is precisely how the adjoint of $\Ex6$ decomposes
under $\SL(2,\bbR)\times\SU(3,3)\subset \Ex6$. 

Since the full orbit for $\JJ_a$ corresponded to a $\ee^{C^-}\ee^{\tB}
\ee^{-B}$ transformation on $\JJ^{(0)}_a$ we might expect the same for
$\lambda$. That is, the generic element is given by 
\begin{equation}
\label{Clambda}
   \lambda = \ee^{C^-}\ee^{\tB} \ee^{-B} \lambda^{(0)} \ .
\end{equation}
However, this is not yet the full story. Such transformations do not
quite fill out the orbit. Instead, 12 degrees for freedom are
missing. We will come back to this point in the following section.  

As we have mentioned, it is well-known~\cite{MS,cecotti,dWVP} that
the space~\eqref{vec-cone} admits a special K\"ahler metric. We now
turn to the explicit construction of this metric, which will follow
exactly Hitchin's construction of the corresponding special K\"ahler
metric on the space $O(6,6)/\SU(3,3)\times\bbR^+$ of $\re\Phi^\pm$ \cite{Hitchin}. As
there, we start with a natural symplectic structure, since, as
discussed in section~\ref{app:Tsub}, by definition $\Es7$ preserves a
symplectic structure $\mathcal{S}(\lambda,\lambda')$ on the
fundamental representation. The complex structure, and hence special
K\"ahler geometry then arise from the natural generalisation of the
Hitchin function~\eqref{KrigidH}. Instead of the $O(6,6)$ spinor
quartic invariant we take the quartic invariant $Q(\lambda)$ in (\ref{invs}) that
defines the $\Es7$ group.
We then define the Hitchin function
\begin{equation}
   H(\lambda) = \sqrt{Q(\lambda)} \ . 
\end{equation}
As before, one can view $H(\lambda)$ as a Hamiltonian and define the
corresponding Hamiltonian vector field $\hat{\lambda}$, given by, for
any $\nu$ in the $\rep{56}$ representation, 
\begin{equation}
   \mathcal{S}(\nu,\hat{\lambda}) 
      = - \nu^\Ea\, \frac{\der H}{\der\lambda^\Ea} \ , 
\end{equation}
where $\Ea=1,\dots,56$ runs over the elements of the fundamental
representation. Explicitly we have
\begin{equation}
\begin{aligned}
   \hat{\lambda}^{i\Ta} 
       &= \frac{1}{2H}\mukai{\lambda^+}{\Gamma^\Ta{}_\Tb\lambda^+}
             \lambda^{i\Tb}
          - \frac{1}{H}\left(\lambda^i\cdot\lambda_j\right)
             \lambda^{j\Ta}, \\
   \hat{\lambda}^+ 
       &= - \frac{1}{2H}\left(
          \frac{1}{12}\mukai{\lambda^+}{\Gamma_{\Ta\Tb}\lambda^+}
          - \epsilon_{ij}\lambda^i_\Ta\lambda^j_\Tb 
          \right) \Gamma^{\Ta\Tb}\lambda^+ \ .  
\end{aligned}
\end{equation}
The complex structure on $\Mrigid^+$ is then given by 
\begin{equation}
\label{JSK}
   {\JSK}^\Ea{}_\Eb
     = \frac{\der\hat{\lambda}^\Ea}{\der\lambda^\Eb} \ . 
\end{equation}
Equivalently the metric on $\Mrigid^+$ is given by the Hessian
\begin{equation}
   \gSK_{\Ea\Eb} 
      = \frac{\der H}{\der\lambda^\Ea\der\lambda^\Eb} \ . 
\end{equation}
Following the same arguments of~\cite{Hitchin} it is easy to show that
this metric is special K\"ahler. Finally, note that one can define the
holomorphic object, analogous to $\Phi^+$, 
\begin{equation}\label{Ldef}
   \LL = \tfrac{1}{2}\big( \lambda + \ii\hat{\lambda} \big) \ , 
\end{equation}
such that the K\"ahler potential is given by 
\begin{equation}
   \ee^{-\KSK} = H(\lambda) = \ii\mathcal{S}(\LL,\bar{\LL}) \ . 
\end{equation}
On the subspace $\ee^{-B}\lambda^{(0)}=(0,\rePhi^+)$ we have 
\begin{equation}
\label{Lsub}
\begin{aligned}
   H(\lambda) 
      &= \sqrt{q(\rePhi^+)} \ , \\
   \LL &= (0,\Phi^+) \ , \\
   \JSK\cdot\nu 
      &= \big( \mathcal{J}^{-\Ta}{}_\Tb\nu^{i\Tb},
         \Jhit^+\cdot\nu^+ \big) \ , 
\end{aligned}   
\end{equation}
where $q(\rePhi^+)$ is the spinor quartic invariant~\eqref{Qdef} and
$\Jhit^+$ is the Hitchin complex structure~\eqref{Jhit} on the spinor
space. Thus we see that the special K\"ahler metric on
$\bbR^+\times\Es7/\Ex6$ reduces to the special K\"ahler metric on
$\bbR^+\times O(6,6)/\SU(3,3)$ on this subspace.


\subsection{Hypermultiplets and vector multiplets: compatibility
  conditions and $\SU(8)$ representations} 
\label{compat}

In this section we turn to the question of compatibility between the
structures arising in the vector multiplet and hypermultiplet sectors.  

To start with, note that the vector multiplet moduli space $\MSK^+$ is
54-dimensional, whereas the original 
$O(6,6)/U(3,3)$ space was 30-dimensional, and one would expect no
additional degrees of freedom in this sector. 
A partial answer to this discrepancy is that, as in the $O(6,6)$ case,
we expect there to be some compatibility condition between the
hypermultiplet $\SO^*(12)$ structure and the vector multiplet $\Ex6$
structure. This can be seen by considering the way the supergravity
degrees of freedom are encoded in $\Es7$.  Recall that
in the EGG the internal bosonic metric, form-field and axion-dilaton
degrees of freedom are encoded in the exceptional generalised
metric~\eqref{EGM}. This defines an $\hg$ structure on the exceptional
generalised tangent space $E$. The fermions in the supergravity
transform under the local $\SU(8)$ group, as do the supersymmetry
parameters. In particular, recall that for the
$O(6,6)$ case the pair $(\eta^1,\eta^2)$ transforms under
$\Spin(6)\times\Spin(6)\simeq\SU(4)\times\SU(4)$. Thus to see the
$\SU(8)$ transformation properties we simply rewrite our original
spinor decomposition~\eqref{decompepsilon} as
\begin{equation}
\label{theta-decomp}
   \begin{pmatrix} \epsilon^1 \\ \epsilon^2 \end{pmatrix}
      = \varepsilon^1_+ \otimes (\theta^1)^*
         + \varepsilon^2_+ \otimes (\theta^2)^* + \text{c.c.} 
\end{equation}
where $(\theta^I)^*$ are the complex conjugates of two
elements $\theta^1$ and $\theta^2$ of the $\rep{8}$ representation of
$\SU(8)$  
\begin{equation}
\label{thetas}
   \theta^1 = \begin{pmatrix} \eta^1_+ \\ 0  \end{pmatrix} \ , \qquad
   \theta^2 = \begin{pmatrix} 0 \\ \eta^2_-  \end{pmatrix} \ . 
\end{equation}
Together, the pair $(\theta^1,\theta^2)$ is invariant under
$\SU(6)\subset\SU(8)$ transformations. Thus 
we see that for $N=2$ supersymmetry, comparing the generalised and
exceptional generalised geometries we have the structures 
\begin{equation}
\begin{aligned}
   \text{gen. geom.:} \quad &
   \SU(3)\times\SU(3) && \subset O(6)\times O(6) && \subset O(6,6) \\
   \text{exceptional gen. geom.:} \quad &
   \SU(6) && \subset \hg && \subset \Es7 \ . 
\end{aligned}
\end{equation}
Thus in general we expect the hypermultiplet $\SO^*(12)$ structure and
the vector multiplet $\Ex6$ structure to be constrained such that they
have a common $\SU(6)$ subgroup, that is, as embedding in $\Es7$ 
compatibility requires
\begin{equation}
\label{consist}
   \SO^*(12)\cap\Ex6=\SU(6) \ .
\end{equation}
Thus together the consistent hypermultiplet and vector multiplet
moduli spaces, coming from the cones $\MHK$ and $\Mrigid^+$, describe
the coset space  
\begin{equation}
   \Mfull = \frac{\Es7}{\SU(6)}\times\bbR^+\times\bbR^+ \ ,
\end{equation}
or if we go to $\MQK$ and $\MSK^+$, we have $\Es7/(\SU(6)\times
U(2))$. The $U(2)$ factor corresponds to R-symmetry rotations on the
two $\theta^1$ and $\theta^2$. This last space is 94-dimensional with
signature $(70,24)$, while the sum of the dimensions of $\MQK$ and $\MSK$ 
is 118. This implies  that the compatibility condition~\eqref{consist}
should impose 24 conditions. 
Let us see if this is indeed the case:
requiring that the $SO^*(12)$ stabiliser of $K_a$ shares a common $SU(6)$ subgroup
with the $\Ex6$ stabiliser of $\lambda$  translates into the requirement
\begin{equation}
\label{JLcomp}
   \JJ_a \cdot \lambda = 0\ , \quad a=1,2,3 \ , 
\end{equation}
where we are simply taking the adjoint action on the fundamental
representation. It is equivalent to $\JJ_+\cdot\LL=0$ (with $L$
defined in \eqref{Ldef}). In particular
we see from \eqref{adjfund} and \eqref{JE7} that 
\begin{equation}
\begin{aligned}
   \big(\JJ^{(0)}_+\cdot L^{(0)} \big)^{iA}
      &= u^i \mukai{\Phi^-}{\Gamma^A\Phi^+} \ , \\
   \big(\JJ^{(0)}_+\cdot L^{(0)} \big)^+ 
      &= 0 \ , 
\end{aligned}
\end{equation}
so, at this point, compatibility is equivalent to the compatibility
condition~\eqref{comp1} between $\Phi^+$ and $\Phi^-$, which amounts
only to 12 conditions. Thus there are 12 conditions unaccounted for.
On the other hand, if we count up the degrees of freedom
in $\Phi^+,\Phi^-$, $C^-$ and $\phi,\tilde B$ we get
$48+32+2=82$, while $\Mfull$ is 94-dimensional, again leaving 12 
degrees of freedom unaccounted for. 

Looking at our spinor ansatz~\eqref{thetas} we can immediately see
what is missing: the expressions for $\theta^1$ and $\theta^2$ are not
generic. A generic $\SU(6)$ structure is given by 
\begin{equation}
\label{theta-gen}
   \theta^1 
      = \begin{pmatrix} \eta^1_+ \\ \tilde{\eta}^1_- \end{pmatrix} \ , 
      \qquad
   \theta^2 
      = \begin{pmatrix} \tilde{\eta}^2_+ \\ \eta^2_-  \end{pmatrix} \ . 
\end{equation}
$\tilde \eta^1, \tilde \eta^2$ introduce 16 real new parameters. However, there is a $U(2)$ R-symmetry rotating the two $\theta^I$, which can be used to remove four parameters, and therefore there are indeed precisely 12 new degrees of freedom. 
The special property of the
ansatz~\eqref{thetas} is that the $\SU(6)$ structure decomposes into
$\SU(3)\times\SU(3)$ under $O(6,6)\subset\Es7$. There is nothing about
this freedom that is special to the $\Es7$ formulation, we could
always have used it to generalise our original $N=2$
ansatz~\eqref{decompepsilon} even in the $O(6,6)$ formulation. In
terms of the $\SU(6)$ subgroup, these extra degrees of freedom
transform in the $\rep{6}+\bar{\rep{6}}$ representation. Including
these degrees of freedom, as we have mentioned, the ``local'' version of
$\Mfull$, i.e.~the coset
$\Es7/(\SU(6)\times U(2))$ contains 24 non-physical modes beyond the
70 parameterised by the supergravity degrees of freedom $g$, $B$, $C^-$
and $(\phi,\tB)$. As before, we expect that these are related
to the massive spin-$\frac32$ degrees of freedom and can be gauged
away. Alternatively we can view this as simply projecting out all the
$\rep{6}$ and $\bar{\rep{6}}$ degrees of freedom. 

Regarding the number of conditions imposed by compatibility,
we saw in the previous section that the generic element  
$\lambda=e^{C^-} \lambda^{(0)}$ has 44 degrees of freedom and 
therefore the action of $C^-$ does not fill out the full 56-dimensional orbit
$\Mrigid^+$. The missing 
12 extra degrees of freedom are  precisely
those that correspond to using the generic
spinor ansatz~\eqref{theta-gen}. Thus we can write the generic element
$\lambda$ in the form~\eqref{Clambda}, 
provided
$\lambda^{(0)}=\tilde{g}(0,\rePhi^+)$ where
$\tilde{g}\in\SU(6)\subset\hg\subset\Es7$ is the element which
transforms the restricted ansatz~\eqref{thetas} to the general
form~\eqref{theta-gen}. Equivalently we can write $\lambda$ in the
terms of the $\hg$ decomposition~\eqref{eq:hg-decomp} of $\Es7$, as 
\begin{equation}
\label{lambda-spinor}
\begin{gathered}
   \lambda = (\lambda^{\alpha\beta},\lambda_{\alpha\beta})
      = \ee^{C^-}\ee^{\tB}\ee^{-B} \lambda^{(0)} \ , \\
   \lambda^{(0)\alpha\beta} 
       ´= \epsilon_{IJ}\theta^{I\alpha}\theta^{J\beta} \ , \qquad
   \bar{\lambda}^{(0)}_{\alpha\beta} 
       = \epsilon_{IJ} \theta^{I\ast}_\alpha\theta^{J\ast}_\beta \ , 
\end{gathered}
\end{equation}
where $\epsilon_{12}=-\epsilon_{21}=1$ and $\theta^I$ are the generic
spinors~\eqref{theta-gen}. In this form, it is easy to see that
$\lambda$ is invariant under $\SU(2)_\textrm{R}$. For a generic element in the $\rep{56}$, compatibility requirement (\ref{JLcomp}) 
amounts to 24 conditions, as opposed to 12 for the case of $\lambda$ belonging to the 44-dimensional
orbit.

From a supergravity perspective the appearance of $C^-$ in the vector
multiplet sector is unexpected, since in simple Calabi--Yau models it
corresponds to a hypermultiplet degree of freedom. Equally odd is that
the moduli space $\Oo(6,6)/\SU(3,3)\times\bbR^+$ spanned by $\Phi^+$
has been promoted to the larger space~\eqref{vec-cone}. As we have
discussed these extra degrees of freedom can be accounted for by
compatibility condition~\eqref{JLcomp} and the generalised spinor
ansatz~\eqref{theta-gen}. It is helpful to note, however, that if we
project out all the $\SU(3)\times\SU(3)$ triplets (or equivalently the
$\SU(6)$ representations $\rep{6}$ and $\rep{\bar{6}}$), then the moduli
spaces of $\lambda$ and $\Phi^+$ agree, and $C^-$ does not appear in
the vector multiplet sector. As we argued, projecting out the triplets precisely ensures the
absence of additional (massive) spin-$\frac{3}{2}$ gravitino
multiplets. Their presence would change the standard form of the $N=2$
supergravity and in particular the decoupling of vector multiplets and
hypermultiplets would no longer hold. The construction presented
here shows that if we include all representations then we can rewrite
the field space in terms of $\Es7$ objects, but with vector multiplet
and hypermultiplet moduli spaces coupled through the compatibility
condition~\eqref{JLcomp}.
The expectation is that the additional (non-physical) coupled
degrees of freedom are associated to the massive spin-$\frac32$
multiplets and can be gauged away. 

In the previous sections, we actually always used the restricted
$\SU(3)\times\SU(3)$ spinor ansatz given by~\eqref{thetas} when making
explicit calculations. This means that we do not quite fill out
the true moduli spaces. Nonetheless our final expressions are
written as though all the structures were generic, so that, although
we calculated in a slightly restricted case, we believe the resulting
formulas are in fact true in general. One key point, as we will see, is
that the ansatz~\eqref{thetas} is general enough to encode the generic 
supersymmetric $N=1$ vacua. 

Let us end this section by noting that the coset space $\MQK$ describing
the hypermultiplet moduli space can also be
simply described in terms of the $\hg$ decomposition of $\Es7$
directly as bilinears of the $\SU(8)$ spinors $\theta^I$. Using the
notation of~\eqref{eq:hg-decomp} we have 
\begin{equation}
\label{J-spinor}
\begin{gathered}
   \JJ_a = (\JJ_a{}^\beta{}_\beta,\JJ_a^{\alpha\beta\gamma\delta},
          \bar\JJ_{a\,\alpha\beta\gamma\delta} ) 
      = \ee^{C^-}\,\ee^{\tB}\,\ee^{-B}\, \ee^{-\phi} \JJ^{(0)}_a \\
   \JJ^{(0)}_a{}^\alpha{}_\beta 
      = \tfrac12  \sigma_{a\,I}{}^{J} \theta^{I\,\alpha} 
         \bar{\theta}_{J\,\beta} \ , \qquad
   \JJ_a^{(0)\alpha\beta\gamma\delta} = 0 \ , \qquad
   \bar{\JJ}^{(0)}_{a\,\alpha\beta\gamma\delta} = 0 \ , 
\end{gathered}
\end{equation}
where $\sigma_a$ are the Pauli matrices, $\ee^{C^-}$ is the
action of the RR scalars $C^-$ in $\Es7$ as above, $\ee^{\tB}$ is the
axion action in $\SL(2,\bbR)\subset\Es7$ and $\ee^{-B}$ is just the
usual action of the NS $B$-field in $O(6,6)$ embedded in $\Es7$. Here
we span the full 63-dimensional subspace just using the restricted
ansatz (\ref{thetas}). This reproduces~\eqref{JE7} in the
gauge~\eqref{ident}. From~\eqref{theta-decomp} we see that the
$\SU(2)_\textrm{R}$ R-symmetry acts on the doublet $\theta^I$. Given
the form~\eqref{J-spinor}, this translates into rotations of the
triplet $\JJ_a$ as expected. It is also easy to check compatibility
with $\lambda$, namely  using~\eqref{lambda-spinor}
and~\eqref{spinorE7} we see that $\JJ_a\cdot\lambda=0$.


\section{Killing prepotentials and $N=1$ vacua}


Thus far we have identified how the vector multiplet and
hypermultiplet degrees of freedom are naturally encoded as orbits in
the fundamental and adjoint representations of $\Es7$ respectively. In
particular we have identified the corresponding special K\"ahler and
quaternionic geometries which govern the kinetic terms of the fields. 

In this section we turn first to the Killing prepotential terms in
the $N=2$ action, which encode the gauging of the vector multiplets
and the scalar potential, and second, we briefly discuss the form of the
$N=1$ supersymmetric vacua equations in this formulation. Both
objects, unlike the kinetic terms, now depend on the differential
structure of the EGG, but, as we will see, can still be written in
$\Es7$ form. 

\subsection{Killing prepotentials}
\label{E7P}

Since we are interested in the differential structure, we start by
introducing an embedding of the exterior derivative $\dd$ in $\Es7$. 
Taking a slightly different version of the EGT, the one weighted
by $(\Lambda^6T^*M)^{-1/2}$ (see (\ref{Ep}), \eqref{EGT}), namely 
\begin{equation}
\label{Em12}
   E_{-1/2} = TM \oplus T^*M 
       \oplus \Lambda^5TM 
       \oplus \left(TM\otimes\Lambda^6TM\right) 
       \oplus \Lambda^\textrm{even}TM \ ,
\end{equation}
we then embed the exterior derivative $\dd$ in the one-form
component $T^*M$. In the notation of Eq. (\ref{56}) this defines an
element of the $\rep{56}$
\begin{equation}
\label{Ddef}
   D = ( D^{iA}, D^+ ) = ( v^i\dd^A, 0 )\ ,\qquad A=1,\ldots,12\ ,
\end{equation}
where the operator $\dd^A$ only has entries in its `lower' six
components, i.e.~$\dd^A=(0,\der_m)$ where $m=1,\ldots,6$. 

The form of the $N=2$ prepotentials~\eqref{Pdef} and~\eqref{PHKC}
suggests that the Killing prepotentials on the hyperk\"ahler cone can be
written in terms of $\Es7$ objects as  
\beq
\label{KillingE7}
   \cP^\textrm{HKC}_a =  \ii \cS(\LL, D\JJ_a)\ ,
\eeq
where the symplectic pairing $\cS$ is defined in \eqref{invs} and
$D\JJ_a$ represents the $\rep{56}$ component in the product
$\rep{56}\times\rep{133}$, that is, the usual action of the adjoint on
the fundamental representation. Let us show that this is indeed the
case. We will give the proof for the slightly restricted
ansatz~\eqref{thetas}. However, given it can be put in $\Es7$ form, we believe
it to be true in general.

We first note that the compatibility conditions~\eqref{JLcomp} imply
that an $\SU(2)$ rotation on $\JJ_a$ simply rotates the prepotentials
$\mathcal{P}_a$ as expected. In particular the terms with derivatives
of the rotation matrix drop out. Thus in calculating~\eqref{KillingE7}
we can effectively take $\JJ^{(0)\prime}_a=\JJ^{(0)}_a$
in~\eqref{J-angle}, and hence
$\JJ_a=\ee^{C-}\ee^{\tB}\ee^{-B}\JJ^{(0)}_a$. We also 
have $\LL=\ee^{C^-}\ee^{\tB}\ee^{-B}\LL_0$ so we can rewrite
$\mathcal{S}(\LL,D\JJ_a)
=\mathcal{S}(\ee^{\tB}\ee^{-B}\LL_0,\ee^{-C^-}D\JJ_a)$. The 
explicit calculation of $\ee^{-C^-}D\JJ_a$ is given in 
appendix~\ref{Killing}. Using $\ee^{\tB}\ee^{-B}\LL_0=(0,\Phi^+)$ we
then find 
\beq
\label{KillingE72}
\begin{aligned}
 \cP^\textrm{HKC}_+ &= \ii \cS(\LL, D\JJ_+)   
      =  \ii (uv) \mukai{\Phi^+}{\dd\Phi^-} \ , \\
 \cP^\textrm{HKC}_- &=  \ii \cS(\LL, D\JJ_-)   
      = \ii (\bar{u}v) \mukai{\Phi^+}{\dd\bar{\Phi}^-} \ , \\
 \cP^\textrm{HKC}_3 &=  \ii \cS(\LL, D\JJ_3)   
      = - \ii \frac{(uv)(\bar{u}v)}{\sqrt{-2\ii u\bar{u}}}\,
        \sqrt{\ii\mukai{\Phi^-}{\bar \Phi^-}}\, \mukai{\Phi^+}{\dd C^-} \ .
\end{aligned}
\eeq
The next step is to compare these expressions with~\eqref{PHKC}
and~\eqref{Pdef}. We find 
\begin{equation}\label{uv1}
   uv = -2\ee^{\frac{1}{2}K^-+\phi} \chi \, \ee^{\ii\alpha} \ ,
   \qquad
   \frac{(uv)(\bar{u}v)}{\sqrt{-2\ii u\bar{u}}}
        \sqrt{\ii\mukai{\Phi^-}{\bar \Phi^-}} = \ee^{2\phi} \chi \ .
\end{equation}
Inserting the first equation into the second and using
\eqref{Kahlerpot}, we confirm Eq.~\eqref{chiWolf} for the the
hyperk\"ahler potential, and thus are left with 
\begin{equation}\label{uv2}
   uv = - \sqrt{-\tfrac12 \ii u\bar u} \, \ee^{\phi} \ee^{\ii\alpha}\ .
\end{equation}
We have already argued that the conformal compensator degrees of freedom $Y$
correspond to a common rescaling of $u^i$. Using a parameterisation adapted to the convention of appendix~\ref{app:EGG} where we take $v^1=1$ and $v^2=0$, we write \footnote{\label{foot:covu}Introducing an orthogonal vector $\omega^i$ (such that $v\omega=1$)
we can also write more covariantly, $u=Y(S v+\omega)$.}
\begin{equation}\label{Ydef}
\begin{pmatrix} u^1\\ u^2 \end{pmatrix} 
   = Y \begin{pmatrix} S\\ - 1 \end{pmatrix}\ .
\end{equation}
 Since $u^i$ transforms as an $\SLR$ doublet according to
\begin{equation}
\begin{pmatrix} u^1\\ u^2 \end{pmatrix} \to
\begin{pmatrix} 
      a & b\\ c & d
      \end{pmatrix}
\begin{pmatrix} u^1\\ u^2 \end{pmatrix} \ , \qquad ad-bc =1\ ,
\end{equation}
one checks that $S$ indeed transforms by fractional linear transformations
\begin{equation}
S \to - \frac{aS-b}{cS-d}\ .
\end{equation}
Inserting \eqref{Ydef} (or its covariant version, as in footnote
\ref{foot:covu}) into \eqref{uv2} yields  
\begin{equation}
\label{uresult}
   Y = r\ee^{\ii\alpha} = - uv \  , 
   \qquad 
   S-\bar S = 2\ii\ee^{-2\phi}\ .
\end{equation}
As anticipated the comparison of the Killing prepotentials successfully
determined the dilaton dependence in $u^i$. 
Inserting back \eqref{uresult} into \eqref{chiWolf} we indeed find the
hyperk\"ahler potential $\chi$ as given in \eqref{chiVandoren}, where, in the
$O(6,6)$ version, one uses the rescaling ambiguity between $u^i$ and $\Phi^-$ mentioned below~\eqref{J-angle} to interpret $\sqrt{r}$ as the overall scale of $\Phi^-$.  
We also note that, as expected we can generate the axion component of
$S$ by the $\tB$-transformation $\SL(2,\bbR)$
action given in~\eqref{potentials}. Explicitly, if we get the map 
\begin{equation}
   u^{(0)} = Y \begin{pmatrix} \ii\ee^{-2\phi} \\ - 1 \end{pmatrix}
   \mapsto (\ee^{\tB})^i_{\hp{i}j} u^{(0) \, j}
      = Y \begin{pmatrix} \ii\ee^{-2\phi} \\ - 1 \end{pmatrix}
         + Y \begin{pmatrix} \tB \\ - 1 \end{pmatrix}
      = Y \begin{pmatrix} S \\ - 1 \end{pmatrix} \ , 
\end{equation}
where $S=\tB+\ii\ee^{-2\phi}$. 


\subsection{$N=1$ vacuum equations}
\label{N=1text}

We now turn to a brief discussion of how the equations defining the $N=1$
vacua for type II compactifications~\cite{GMPT2} can be reformulated
in $\Es7$ language. We will simply give a sketch of the form of the
corresponding equations leaving a full analysis for future work. 
Specifically we give $\Es7$ expressions which encode the $N=1$ equations in their 
$O(6,6)$ spinor components, and then discuss to what extent these hold in general. 

Recall that defining an $N=1$ vacuum picks out a particular
preserved supersymmetry in the $N=2$ effective theory breaking the
$\SU(2)_R$ symmetry to $U(1)_R$. Correspondingly, as discussed
in~\cite{GLW,GLW2}, one can identify the $N=1$ superpotential $W$ as a
complex linear combination of the $N=2$ Killing prepotentials
$\cP_a$. The remaining orthogonal combination is then related to the
$N=1$ $D$-term. Concretely, given the
decomposition~\eqref{theta-decomp}, one identifies the preserved $N=1$
supersymmetry as $\varepsilon=\bar n^I \varepsilon_I$ where $n_I$ is a
normalised $SU(2)_R$ doublet $\bar n^I$ (satisfying $\bar n^I
n_I=1$). 
Writing $n^1=a$ and $n^2=\bar{b}$, one then has the $U(1)_R$
doublet and singlet combinations
\begin{equation}
\label{wr}
\begin{aligned}
   \ee^{K/2} W &= \ee^{K^+/2} w^a \cP_a , 
   & && &&
   (w^+,w^-,w^3) &= \left(a^2, -\bar b^2, -2a \bar b\right) , \\
   \mathcal{D} &= r^a \cP_a . 
   & && &&
   (r^+,r^-,r^3) &= \left(ab, \bar a \bar b , |a|^2-|b|^2 \right) , 
\end{aligned}  
\end{equation}
corresponding to the superpotential and $D$-term respectively.\footnote{\label{Mmatrix}
This is equivalent to making an $\SU(2)_R$ rotation by the matrix
\beq
M^I{}_J= \begin{pmatrix}  a &  \bar b \\ - b &  \bar a \end{pmatrix} \ ,
 \qquad |a|^2+|b|^2=1 \ . \nn
\eeq
}

In general the equations governing $N=1$ vacua \cite{GMPT2} should be
obtainable by extremising the superpotential and setting the
D-term to zero~\cite{MK,BC}. Using the parameterisation above, the preserved 
supersymmetries take the form 
\begin{equation}
\label{N=1SU8spinor}
   \epsilon = \varepsilon_+ \otimes \ee^{A/2}\,\theta^* + \text{c.c.} \ , 
   	\qquad \text{where} \qquad
   \theta =  \begin{pmatrix} a\eta_+^1 \\ \bar{b}\eta_-^2 \end{pmatrix} \ , 
\end{equation}
and $A$ is the warp factor in front of the four-dimensional metric. In the special case of $|a|^2=|b|^2$, and for zero cosmological constant on the four-dimensional
space, the corresponding equations, in our conventions, are~\cite{MK,toma} 
\begin{equation}
\label{eq:N=1vac}
\begin{aligned}
   \dd \left( \ee^{3A-\phiten}\Phi^{\prime+} \right) &= 0 \ , \\
   \left.\left[ \dd \left( \ee^{-\phiten}\re\Phi^{\prime-}\right) 
      - \ii G\right]\right|_{1,0} &= 0 \ , \\
   \dd \left( \ee^{2A-\phiten}\im\Phi^{\prime-} \right) &= 0 \ . 
\end{aligned}
\end{equation}
where 
\begin{equation}
\label{Phiprime}
   \Phi^{\prime+} = a\bar{b} \Phi^+ \ , \qquad 
   \Phi^{\prime-} = ab \Phi^- \ ,
\end{equation}
while $\phiten$ is the ten-dimensional dilaton and $|_{1,0}$ represents the
projection onto the $+\ii$ eigenspace using the Hitchin complex
structure $\Jhit^+$. 
From (\ref{wr}) it is not hard to see that the first two equations in
(\ref{eq:N=1vac}) can essentially be obtained from variations of the
superpotential while the third one corresponds to the D-term equation.  

An important point here is that although our original parameterisation
of the $N=2$ supersymmetries was not completely generic, the
$N=1$ vacuum parameterisation is generic. Recall that the $N=2$
$\SU(8)$ spinor ansatz given in~\eqref{theta-decomp} is restricted
since some components vanish. Nonetheless, the $N=1$ 
supersymmetry~\eqref{N=1SU8spinor} can be written with this ansatz as $a\theta^1+\bar{b}
\theta^2$ and is completely generic. The stabiliser of $\theta$ is
$\SU(7)$. Thus, in terms of structures, since the $N=1$
vacuum is determined solely by $\theta$, it defines a particular
$\SU(7)$ structure on the exceptional generalised tangent space
$E$. By contrast, as we have seen, to define the $N=2$ effective
theory, we require two spinors $\theta^1$ and $\theta^2$ defining a
$\SU(6)$ structure.

In the language of $\Es7$,  the preserved $N=1$ symmetry picks a particular
complex structure on the hyperk\"ahler cone,  and uses it to 
define the $N=1$ chiral field inside the hypermultiplet. In terms of $\omega^a$, $r^a$, 
this reads\footnote{
One can also derive these expressions from \eqref{J-spinor}, rotating the $\SU(8)$ spinors $\theta^I$ by the matrix $M$ given in footnote~\ref{Mmatrix}.}
\begin{equation}
      \JJ_3^{{(0)}\prime} = r^a \JJ^{(0)}_a \ , \qquad \JJ_+^{{(0)}\prime} = w^a \JJ^{(0)}_a \ ,
\end{equation}
where 
$\JJ_3^{\prime}$ corresponds to the particular $N=1$ complex
structure and $\JJ_+^{\prime}$ defines the chiral field. 
Parameterising $a$ and $b$ as 
\begin{equation}
   a = \sin\tfrac{1}{2}\theta \ee^{\ii \gamma} \ , \qquad 
   b =  \cos\tfrac{1}{2}\theta \ee^{\ii \beta} \ ,
\end{equation}
and defining $\psi\equiv \gamma - \beta,  \alpha\equiv \gamma+\beta$, we get $r^a
\JJ^{(0)}_a=\JJ_3^{{(0)}'}$ then matches the expressions~\eqref{J-angle}. As mentioned before, 
$\gamma+\beta$  can be identified with the $\U(1)_R$ angle
$\alpha$ in the compensator $Y$  (see~eq.~\ref{uresult}), while
$\theta$ and $\gamma-\beta$ are the Euler angles. From the expression~\eqref{Phiprime} 
for $\Phi^{\prime+}$ we see that there should be a rescaling of $\LL$ by $a\bar{b}$. Given 
the $\SU(8)$ covariant expression~\eqref{lambda-spinor} we see that the phase 
of this rescaling corresponds to diagonal $U(1)$ in the $U(2)_R$ symmetry given by 
$\theta^I\mapsto\ee^{\ii(\gamma-\beta)/2}\theta^I$. 

For the case $|a|^2=|b|^2=1$ corresponding to $\theta=-\pi/2$, given we can always absorb two phases by redefining $\eta^1$ and $\eta^2$, without loss of generality we can set 
$\gamma+\beta=\alpha=\pi/2$, and $\psi=\pi/2$, so that  
\begin{equation}
\label{rotation}
   \JJ^{{(0)}\prime}_3 =  \JJ_2^{(0)} \ , \qquad
   \JJ^{{(0)}\prime}_+ = - \JJ_1^{(0)} + \ii \JJ_3^{(0)} \ .
\end{equation}
By comparing the expressions~\eqref{JE7} and~\eqref{Lsub} for
$\JJ_a^{(0)\prime}$ and $\LL$ one is led the
naive conjecture for the generic $\Es7$ form of the $N=1$ equations
\begin{align}
\label{Leqn}
   D\left(\ee^{3A-\phiten}\LL\right) &= 0 \ , \\
\label{Jeqn}
   \left.D\JJ_+\right|_{1,0} &= 0 \ , \\
\label{Deqn}
   D\left(\ee^{2A}\JJ_3\right) &= 0 \ . 
\end{align}
Here $\JJ_a=\ee^{C^-}\ee^{-B}\JJ^{(0)\prime}_a$ and the projector $|_{1,0}$ now projects onto the $+\ii$ eigenspace of $\JSK$, the complex structure on the $\rep{56}$ representation defined by $\LL$ and given in~\eqref{JSK}. Furthermore, in the first line we are taking the
projection onto the $\rep{133}$ representation of
$D\LL\in\rep{56}\times\rep{56}$. Finally we also choose $Y=\ee^{-\phiten}$
in the expression~\eqref{Ydef} for $u^i$. 

To investigate to what extent these equations hold, we again focus on
the simple $|a|^2=|b|^2$ case. Consider first the equation
\eqref{Leqn} for
$\LL$. We need the projection onto the $\rep{133}$
representation of the symmetric $\rep{56}\times\rep{56}$ tensor
product. It is given by  
\begin{equation}
\label{ffa}
\begin{aligned}
   \left(\lambda\cdot\lambda'\right)^i_{\hp{i}j}
      &= 2\epsilon_{jk}(\lambda^{i}\cdot\lambda'^{k}) \ ,\\
   \left(\lambda\cdot\lambda'\right)^A_{\hp{A}B}
      &= 2\epsilon_{ij}\bigl[(\lambda^{iA}\lambda'^j_{\hp{j}B})
         + (\lambda'^{iA}\lambda^j_{\hp{j}B})\bigr]
         + \mukai{\lambda^+}{\Gamma^A_{\hp{A}B}\lambda'^+} \ , \\
  \left(\lambda\cdot\lambda'\right)^{i-}
      &= \left(\lambda^{iA}\Gamma_A\lambda'^+
         + \lambda'^{iA}\Gamma_A\lambda^+\right) \ . 
\end{aligned}
\end{equation}
Writing $\LL=\ee^{C^-}\ee^{-B}\LL_0=\ee^{C^-}(0,\Phi^+)$, using the
same generalised connection as in appendix~\ref{Killing} one finds
\begin{equation}
\begin{aligned}
   \ee^{-C^-}D\left(\ee^{3A-\phiten}\LL\right)^i_{\hp{i}j}
      &=\sqrt{2}\ee^{3A-\phiten}v^iv_j\mukai{\Phi^+}{G} \ , \\
   \ee^{-C^-}D\left(\ee^{3A-\phiten}\LL\right)^\Ta_{\hp{i\Ta}\Tb} 
      &= 0 \ , \\
   \ee^{-C^-}D\left(\ee^{3A-\phiten}\LL\right)^{i-} 
      &= v^i\dd\left(\ee^{3A-\phiten}\Phi^+\right) \ ,
\end{aligned}
\end{equation}
where we have used $G=\sqrt{2}\dd C^-$. We see that the spinor
component $D(\ee^{3A-\phiten}\LL)^{i-}=0$ indeed reproduces the first
equation~\eqref{eq:N=1vac}. The other components vanish provided
$\mukai{\Phi^+}{G}=0$, but it is easy to see that this is implied by the
first equation in (\ref{eq:N=1vac}).

Let us now turn to the hypermultiplet equations. Using the results in
appendix~\ref{Killing}  and the
relations~\eqref{rotation} we find for the spinor components
\begin{equation}
\begin{aligned}
   \ee^{-C^-} D\left(\ee^{2A}K_3\right)^+
      &= \dd\left[ e^{2A} (uv)\im\Phi^- \right] , \\
   \ee^{-C^-} \left(DK_+\right)^+
      &= -\dd\left[ (uv)\re\Phi^-\right] - \ii
      \frac{(uv)(\bar{u}v)}{\sqrt{-2\ii u\bar{u}}}\,
        \sqrt{\ii\mukai{\Phi^-}{\bar \Phi^-}}\dd C^- 
\end{aligned}
\end{equation}
Given $Y=\ee^{-\phiten}$ we have 
\begin{equation}
   uv = -\ee^{-\phiten} \ , 
   \qquad
   -\ii u\bar{u} = 2\ee^{-2\phiten}\ee^{-2\phi} 
      = 2\ee^{-4\phiten} \vol_6 \ , 
\end{equation}
where we have used~\eqref{eq:phifour}. Since in our conventions, 
$\ii \mukai{\Phi^-}{\bar \Phi^-}=8\vol_6$ (see for
instance~\eqref{Kahlerpotsingle}) we then have 
\begin{equation}
\begin{aligned}
   \ee^{-C^-} D\left(\ee^{2A}K_3 \right)^+
      &=  - \dd\left( \ee^{2A-\phiten}\im\Phi^- \right) , \\
   \ee^{-C^-} \left( DK_+\right)^+
      &= \dd\left( \ee^{-\phiten}\re\Phi^-\right) - \ii G , 
\end{aligned}
\end{equation}
On the subspace
$\LL=\ee^{-B}\LL^{(0)}=(0,\Phi^+)$, by equation~\eqref{Lsub}, we have
for the spinor component of the $\rep{56}$ that
$(\JSK\cdot\nu)^+=\Jhit^+\cdot\nu^+$. Hence we see that the spinor components
of~\eqref{Jeqn} and~\eqref{Deqn} do indeed reproduce the corresponding
equations in~\eqref{eq:N=1vac}. 

In summary, we see that we have reproduced the $N=1$ vacuum equations
in $\Es7$ form. The $\LL$ equation is equivalent to the first equation
in~\eqref{eq:N=1vac}. Decomposing into $\Tsub$ representations we see
that spinor components of the $\JJ_a$ equations~\eqref{Jeqn}
and~\eqref{Deqn} are equivalent the the second and third equations
in~\eqref{eq:N=1vac}. However, it is harder to see how the vector
components of these equations can be implied by supersymmetry. It
seems likely that one would need to take further projections to get a
consistent $\Es7$ form of the equations.


\section{Conclusions}
\label{Conc}

In this paper we studied $N=2$ backgrounds of type II string theory 
from a geometric viewpoint where the U-duality group is manifest.
The formalism we have used, known as extended or exceptional
generalised geometry, has been developed in refs.~\cite{Chris,EGG} and  
is an extension of Hitchin's generalised geometry~\cite{Hitchin}. 
The latter framework ``unifies'' the tangent and cotangent bundles,
such that the T-duality group acting on the degrees of freedom in the
NS-sector is manifest. Incorporating the RR-sector requires a
formalism where the full U-duality group acts on some even larger 
generalised tangent bundle.  

For the case at hand the T-duality group is $\Oo(6,6)$ and the NS
degrees of freedom are most conveniently represented by two pure
spinors $\Phi^\pm$. Each of them is invariant under an $\SU(3,3)$
action and thus parameterises a specific $\Oo(6,6)$ orbit
corresponding (after a quotient by $\bbC^*$) to the special K\"ahler
cosets ${\cal M}_{\rm SK}=\Oo(6,6)/U(3,3)$. Together, given a
compatibility condition, the two spinors define an
$\SU(3)\times\SU(3)$ structure. The two $\SU(3)$ factors correspond to
the invariance groups of the two six-dimensional spinors used to
define the $N=2$ background. Incorporating the RR degrees of freedom
enlarges the T-duality group to the U-duality group $\Es7$ which
contains $\Tsub$ as one of its subgroups. Furthermore, $N=2$ requires
that one of the special K\"ahler cosets is promoted to a quaternionic
K\"ahler manifold by means of the c-map~\cite{CFG}. 

In this paper we showed that the NS and RR degrees of freedom which   
populate $N=2$ hypermultiplets can be embedded into the $\rep{133}$
adjoint representation of $\Es7$ and that they parameterise an $\Es7$
orbit corresponding to the  Wolf-space 
${\cal M}_{\rm QK} = \Es7/(\SO^*(12) \times \SU(2))$ which is indeed
quaternionic-K\"ahler. A point in the orbit is defined by a triplet of
real elements $\JJ_a$ fixing an $\SU(2)$ subgroup of
$\Es7$. The $\SU(2)$ action is the R-symmetry of the $N=2$
theory. We also constructed the corresponding hyperk\"ahler cone and
hyperk\"ahler potential following refs.~\cite{Swann,KS-homo2,KS-Wolf}
and showed agreement with the expressions given in~\cite{RVV,pioline}. 

Similarly, we demonstrated that the degrees of freedom which reside in
$N=2$ vector multiplets can be embedded into a complex element $\LL$
of the $\rep{56}$ fundamental representation of $\Es7$. This
parameterises an $\Es7$ orbit, corresponding, after a $\bbC^*$
quotient, to the special K\"ahler coset ${\cal M}_{\rm
  SK}=\Es7/(\Ex6\times U(1))$. Again the  
$U(1)$ factor is an R-symmetry, which combines with the $\SU(2)$
factor from the hypermultiplets to give $U(2)_\textrm{R}$. The
corresponding K\"ahler potential is given by the logarithm of the
square root of the $\Es7$ quartic invariant in complete analogy with
the appearance of the Hitchin function invariant in the $\Oo(6,6)$ 
formalism~\cite{Hitchin,GLW,GLW2}. We noted that to fill out the full
$\Es7$ orbit in the $\rep{56}$ representation requires generalising
the original $N=2$ spinor ansatz to its most generic
form~\eqref{theta-gen}. These six-dimensional spinors 
transform in the fundamental representation of the local U-duality
group $\SU(8)$. By considering an $N=2$ background one picks out two
$\SU(8)$ spinors defining an $\SU(6)$ structure which decomposes into
$\SU(3)\times\SU(3)$ under the T-duality subgroup. Correspondingly,
there is a compatibility condition between the hypermultiplet $\JJ_a$
and vector multiplet $\LL$ such that together they are indeed
invariant under a $\SU(6)$ subgroup of $\Es7$. 

Finally, the Killing prepotentials (or moment maps) $\cP_a$
which determine the scalar potential and couple the two sectors, can  
also be given in an $\Es7$ language. Unlike the kinetic term moduli
spaces, the prepotentials depend on the differential structure of the
exceptional generalised geometry. Nonetheless they also take a simple
form in terms of $\Es7$ objects which we demonstrated agrees with the
expressions calculated in~\cite{GLW2}. We ended by
considering supersymmetric $N=1$ backgrounds. Without giving a
complete description we showed that there are natural $\Es7$
expressions which encode the supersymmetry conditions. However, there
are indications that these probably require additional projections to
give a fully consistent $\Es7$ form. 

There are a number of natural extensions of this work one could
consider. First is of course to consider other dimensions and numbers of
preserved supersymmetries~\cite{N=4}. The former
corresponds to exceptional generalised geometry with different
$E_{d(d)}$ U-duality groups, while the latter correspond to different 
preserved structures within these groups. It is important to note that
the formulation presented here is not $\Es7$ covariant: in particular
the $\Es7$ symmetry is broken by picking out the particular
$\GL(6,\bbR)$ subgroup which acts on the tangent space. The actual
symmetry group for the supergravity theory is a semidirect product of
this diffeomorphism symmetry group with the gauge potential
transformations given in~\eqref{gauge-transf}. Together these form a
parabolic subgroup of $\Es7$. Our formalism is covariant with respect
to this subgroup. One could also consider more general so-called
``non-geometrical'' backgrounds~\cite{non-geom}, which in some sense
incorporate more of the full U-duality group. From the evidence of the
$O(6,6)$-formulation, one expects the $\Es7$ expressions for the
kinetic and potential terms would hold in this more general
context. Furthermore, there also seems to be an
intriguing relation between the moduli spaces which we
discussed in this paper and the charge orbits and moduli spaces 
of extremal black hole attractor
geometries~\cite{Ferrara:2008hwa}.\footnote{We thank S.~Ferrara for
  drawing our attention to this relation.}

One could also consider what this formulation implies for the
topological string, which describes the vector multiplet sector of
type II supergravity. It has been argued~\cite{top-st} that the
$\Oo(6,6)$ Hitchin functionals are the actions of the corresponding
target space theories. In the $\Es7$ formalism, we have seen
that the kinetic terms are now described by the analogue of a Hitchin
function on $\bbR^+\times\Es7/\Ex6$ given by the square-root of the $\Es7$
quartic invariant. Interestingly, as in the $\Oo(6,6)$ case this space
is also a so-called ``prehomogeneous space'', implying it is an open
orbit in the $\rep{56}$ dimensional representation. It would be very
interesting to see how the one-loop calculations of ref.~\cite{pw} are
encoded in this U-duality context, and how this is connected to the 
extension of the topological string to M-theory. 

\vskip 1cm


\subsection*{Acknowledgements}

This work is supported by DFG -- The German Science Foundation, the
European RTN Programs HPRN-CT-2000-00148, HPRN-CT-2000-00122,
HPRN-CT-2000-00131, MRTN-CT-2004-005104, MRTN-CT-2004-503369.  
M.G.\ is partially
supported by ANR grant BLAN06-3-137168.

We have greatly benefited from conversations and correspondence with
P.\ C\'amara, S.\ Ferrara, B.\ Pioline, M.\ Rocek, F. Saueressig and S. Vandoren. 


\appendix


\section{A $\GL(6,\bbR)$ subgroup of $\Es7$}
\label{app:EGG}

We would like to identify a particular embedding of $\GL(6,\bbR)$ in
$\Tsub\subset\Es7$. This will correspond to the action of
diffeomorphisms on the exceptional tangent space in the EGG. For the
$O(6,6)$ factor, we simply take the embedding of $\GL(6,\bbR)$ that
arises in generalised geometry. If $a\in\GL(6,\bbR)$ acts on vectors
$y\in TM$ as $y\mapsto ay$, then $\GL(6,\bbR)\subset O(6,6)$ acts on
the fundamental $\rep{12}$ representation as, given $y+\xi\in TM\oplus  
T^*M$,  
\begin{equation}
   \begin{pmatrix} y \\ \xi \end{pmatrix} 
       \mapsto \begin{pmatrix} a & 0 \\ 0 & (a^{-1})^T \end{pmatrix}
          \begin{pmatrix} y \\ \xi \end{pmatrix} \ . 
\end{equation}
We also choose to embed $\GL(6,\bbR)$ in the $\SL(2,\bbR)$ factor so
that, an $\SL(2,\bbR)$ doublet $w^i$ transforms as 
\begin{equation}
   \begin{pmatrix} w^1 \\ w^2 \end{pmatrix} 
       \mapsto \begin{pmatrix} 
          (\det a)^{-1/2} & 0 \\ 
          0 & (\det a)^{1/2} \end{pmatrix}
          \begin{pmatrix} w^1 \\ w^2 \end{pmatrix} \ . 
\end{equation}
Putting these two ingredients together implies that elements
$\lambda=(\lambda^{i\Ta},\lambda^+)$ of the $\rep{56}$ representation
of $\Es7$, decomposing under $\GL(6,\bbR)$, transform as sections of a
bundle  
\begin{equation}
\begin{aligned}
   E_0 &= \left(\Lambda^6T^*M\right)^{-1/2}\otimes\Big[
            TM \oplus T^*M 
            \\ & \qquad \qquad 
            \oplus \Lambda^5T^*M 
            \oplus \left(T^*M\otimes\Lambda^6T^*M\right) 
            \oplus \Lambda^\textrm{even}T^*M \Big] \ , 
\end{aligned}
\end{equation}
where 
\begin{equation}
\begin{aligned}
   \lambda^{1A} &\in \left(\Lambda^6T^*M\right)^{-1/2}\otimes\Big[
            \Lambda^5T^*M 
            \oplus \left(T^*M\otimes\Lambda^6T^*M\right) \Big] \ , \\
   \lambda^{2A} &\in \left(\Lambda^6T^*M\right)^{-1/2}\otimes\Big[
            TM \oplus T^*M \Big] , \\
   \lambda^+ &\in \left(\Lambda^6T^*M\right)^{-1/2}
            \otimes \Lambda^\textrm{even}T^*M \ . 
\end{aligned}
\end{equation}
It will be helpful to also define spaces weighted by a power of
$\Lambda^6T^*M$ so  
\begin{equation}
\label{Ep}
   E_p = (\Lambda^6T^*M)^p \otimes E_0 \ , 
\end{equation}
such that 
\begin{equation}
\label{EGT}
\begin{aligned}
   E\equiv E_{1/2} &= TM \oplus T^*M 
            \oplus \Lambda^5T^*M 
            \oplus \left(T^*M\otimes\Lambda^6T^*M\right) 
            \oplus \Lambda^\textrm{even}T^*M \ , \\
   E_{-1/2} &= TM \oplus T^*M 
            \oplus \Lambda^5TM 
            \oplus \left(TM\otimes\Lambda^6TM\right) 
            \oplus \Lambda^\textrm{even}TM \ . 
\end{aligned}
\end{equation}
Thus we can write a general element of $E$ as   
\begin{equation}
   \lambda = y + \xi + \fb + \km + \lambda^+ \in E \ , 
\end{equation}
where $y\in TM$, $\xi\in T^*M$, $\fb\in\Lambda^5T^*M$,
$\km\in T^*M\otimes\Lambda^6T^*M$ and
$\lambda^+\in\Lambda^\textrm{even}T^*M$ such that the $(\rep{2,12})$
components are $\lambda^{2A}=y^m+\xi_m$ and
$\lambda^{1m}=\fb^m_{1\dots 6}$ (with $\fb^m_{m_1\dots
  m_6}=6\delta^m_{[m_1}\fb_{m_2\dots m_6]}$) and 
$\lambda^1_m=\km_{m,1\dots 6}$. 

One can also make a corresponding decomposition of the adjoint
representation. We find
\begin{equation}
\label{eq:adj6}
\begin{aligned}
   A_0 &= \left(TM\otimes T^*M\right)
            \oplus \Lambda^2TM \oplus \Lambda^2T^*M
            \\ & \qquad \qquad 
            \oplus \bbR \oplus \Lambda^6T^*M \oplus \Lambda^6TM
            \oplus \Lambda^\textrm{odd}T^*M
            \oplus \Lambda^\textrm{odd}TM \ , 
\end{aligned}
\end{equation}
where $\mu=(\mu^i{}_j,\mu^A{}_B,\mu^{i-})\in A_0$ has 
\begin{equation}
\label{muprop}
\begin{gathered}
   \mu^1{}_1 = -\mu^2{}_2 \in \bbR \ , \qquad
   \mu^1{}_2 \in \Lambda^6T^*M \ , \qquad
   \mu^2{}_1 \in \Lambda^6TM \ , \\
   \mu^A{}_B \in \left(TM\otimes T^*M\right)
            \oplus \Lambda^2TM \oplus \Lambda^2T^*M \ , \\
   \mu^{1-} \in \Lambda^\textrm{odd}T^*M \ , \qquad
   \mu^{2-} \in \Lambda^\textrm{odd}TM \ . 
\end{gathered}
\end{equation}
We also define $A_p=(\Lambda^6T^*M)^p\otimes A_0$. 

Note that we can
identify a subgroup of $\Es7$ generated by the forms in
$A_0$. Introducing a vector $v^i$ with $v^1=1$ and $v^2=0$, we can
write them in a more covariant way as 
\begin{equation}
\begin{aligned}
   \mu^i{}_j &= \tB_{1\dots 6} v^i v_j  \ , &&&
      \tB &\in \Lambda^6T^*M \ , &&&&& \\
   \mu^A{}_B &= \begin{pmatrix} 0 & 0 \\ B & 0 \end{pmatrix} \ , &&&
      B &\in \Lambda^2T^*M \ , \\
   \mu^{i-} &= v^i C^- \ , &&&
      C^- &\in \Lambda^{\textrm{odd}}T^*M \ , 
\end{aligned}
\end{equation}
with the sub-algebra in $\es7$
\begin{equation}
\label{form-alg}
   \big[ B + \tB + C^- , B' +\tB' +  C^{-\prime} \big]
      = 2\mukai{C^-}{C^{-\prime}} 
          + B \wedge C^{-\prime} - B' \wedge C^- \ , 
\end{equation}
that is, the commutator corresponds to a transformation with
$\tB''=2\mukai{C^-}{C^{-\prime}}$ and $C^{-\prime\prime}=B \wedge
C^{-\prime} - B' \wedge C^-$. 
Note that this Lie algebra is nilpotent with index four. The
adjoint action of the subalgebra on an element
$\lambda=y+\xi+\fb+\km+\lambda^+$ is given by  
\begin{equation}
\begin{aligned}
   \big(B+\tB+C^-\big)\cdot \lambda 
      &= -i_y B 
          + \big(i_y\tB + \mukai{C^-}{\hat{\jmath}\lambda^+} \big)
          \\ & \qquad \qquad 
          + \big( jB\wedge \fb + j\xi\wedge \tB 
              + \mukai{C^-}{j\lambda^+} \big)
          + B \wedge \lambda^+ \ , 
\end{aligned}
\end{equation}
where we are using the notation that the symbol $j$ denotes the pure
$T^*M$ index of $T^*M\otimes\Lambda^6T^*M$ and the symbol
$\hat{\jmath}$ denotes the $TM$ index of
$TM\otimes\Lambda^6T^*M\simeq\Lambda^5T^*M$. In particular, given for
any one-form $\gamma$, the element
$\mukai{C^-}{\hat{\jmath}\lambda^+}\in\Lambda^5T^*M$ is given by 
\begin{equation}
   \gamma \wedge \mukai{C^-}{\hat{\jmath}\lambda^+} 
      = \mukai{C^-}{\gamma\wedge\lambda^+} \ . 
\end{equation}
while the elements $jB\wedge\fb$ and $\mukai{C^-}{j\lambda^+}$ in
$T^*M\otimes\Lambda^6T^*M$ are given by, for any vector $y^m$, 
\begin{equation}
\begin{aligned}
   y^m\left(jB \wedge \fb\right)_{m,m_1\dots m_6}
      &= \big(i_yB\wedge\fb\big)_{m_1\dots m_6} \ , \\
   y^m\left(j\xi \wedge \tB\right)_{m,m_1\dots m_6}
      &= \big(i_y\xi \big) \tB_{m_1\dots m_6} \ , \\
   y^m \mukai{C^-}{j\lambda^+}_{m,m_1\dots m_6} 
      &= \mukai{C^-}{i_y\lambda^+}_{m_1\dots m_6} \ . 
\end{aligned}
\end{equation}
%


\section{Computing $D\JJ_a$}
\label{Killing}

We would like to calculate the derivative $D\JJ_a$ where
$D\in\rep{56}$ is the embedding of the exterior derivative given
by~\eqref{Ddef} and in the action of $D$ on $\JJ_a$ we project onto
the $\rep{56}$ representation. 

It will be useful to introduce explicit indices for the components of
the $\rep{56}$ and $\rep{133}$ representations. Viewed as elements of
the larger symplectic group $\Symp(56,\bbR)\supset\Es7$ we can write
$D^\Ea$, and  $\JJ_a^{\Ea\Eb}=\JJ_a^{\Eb\Ea}$, where $\Ea, \Eb=1,\dots,56$ . One then has
$(D\JJ_a)^\Ec=\mathcal{S}_{\Ea\Eb}D^\Ea K_a^{\Eb\Ec}$ where 
$\mathcal{S}_{\Ea\Eb}$ is the symplectic structure~\eqref{invs}. Given
$\mu^{\Ea\Eb}\in\rep{133}$ and some $\Es7$ group element $g$ such that
$\mu'=g\mu$ we have
\begin{equation}
\label{eq:connDK}
\begin{aligned}
   (D\mu^\prime)^\Ec 
      &= \mathcal{S}_{\Ea\Eb}D^\Ea\left(
         g^\Eb_{\hp{\Eb}\Eb'} g^\Ec_{\hp{\Ec}\Ec'} \mu^{\Eb'\Ec'}
          \right) \\
      &= g^\Ec_{\hp{\Ec}\Ec'} \, g^{-1\Ea'}_{\hp{-1\Ea'}\Ea} {\mathcal{S}}_{\Ea'\Eb'}\left[
            D^{\Ea}\mu^{\Eb'\Ec'}
              + \mathcal{A}^{\Ea\Eb'}_{\hp{\Ea\Eb}\Eb}\mu^{\Eb\Ec'}
              + \mathcal{A}^{\Ea\Ec'}_{\hp{\Ea\Ed}\Eb}\mu^{\Eb'\Eb}
            \right] \ , 
\end{aligned}
\end{equation}
where we have used
$\mathcal{S}_{\Ea'\Eb'}g^{\Ea'}_{\hp{\Ea'}\Ea} g^{\Eb'}_{\hp{\Eb'}\Eb}
=\mathcal{S}_{\Ea\Eb}$ and have introduced the generalised connection
\begin{equation}
   \mathcal{A}^{\Ea\Eb}_{\hp{\Ea\Eb}\Ec}
      = g^{-1\Eb}_{\hp{-1\Eb}\Ed}\left(
         D^{\Ea}g^\Ed_{\hp{\Ed}\Ec} \right) 
      \in \rep{56}\times\rep{133}\ . 
\end{equation}
We now specialize to the case where $g=\ee^{C-}$. Given
$D=(v^i\dd^A,0)$ and using~\eqref{potentials} and~\eqref{adjac} we
then have $g^{-1\Ea'}_{\hp{-1\Ea'}\Ea}
D^{\Ea}=D^{\Ea'}$.\footnote{Note that more generally all the form
  field transformations leave $D$ invariant, that is,
  $\ee^{-C^-}D=\ee^{-\tB}D=\ee^BD=D$.} Hence the
connection is given by
\begin{equation}
   \big(\ee^{-C^-} D^\Ea \ee^{C^-}\big)^\Eb_{\hp{\Eb}\Ec}
      = D^\Ea  \delta^{\Eb}_{\hp{\Eb}\Ec} + \mathcal{A}^{\Ea\Eb}_{\hp{\Ea\Eb}\Ec} \ . 
\end{equation}
This can then be calculated using a variant of the
Baker--Campbell--Hausdorff formula which, in this context, reads   
\begin{equation}
\label{BCH}
   \ee^{-C^-}\dd^A\ee^{C^-} 
      = \dd^A \cdot \id
         + \dd^A C^-
         + \tfrac{1}{2!}[\dd^A C^-, C^-]
         + \tfrac{1}{3!}[[\dd^A C^-, C^-],C^-]
         + \dots \ .
\end{equation}
This series truncates at second order with the only non-vanishing
component\footnote{We have $v^iv_i=\epsilon^{ij}v_iv_j=0$.}
\begin{equation}\label{dsimp}
   [\dd^A C^-, C^-]^i_{\hp{i}j} = 2v^iv_j\mukai{\dd^A C^-}{C^-} \ .
\end{equation}
Given
$(\ee^{-C^-}D^\Ea\JJ_a)^{\Eb\Ec}=
\big[(\ee^{-C^-}D^\Ea\ee^{C^-})\ee^{-B}\JJ_a^{(0)}\big]^{\Eb\Ec}$ and
using~\eqref{JE7} and the adjoint action of the generalised connection we
find the nonzero components 
\begin{equation}
\begin{aligned}
   \big(\ee^{-C^-}D^{iA}\JJ_+\big)^j_{\hp{j}k}
      &= v^i\mukai{\dd^AC^-}{\Phi^-} \left(v^ju_k+u^jv_k\right) \ , \\
   \big(\ee^{-C^-}D^{iA}\JJ_+\big)^B_{\hp{B}C}
      &= v^i(uv)\mukai{\dd^AC} {\Gamma^B_{\hp{B}C}\Phi^-} \ , \\
   \big(\ee^{-C^-}D^{iA}\JJ_+\big)^{j-}
      &= v^i\dd^A\left(u^j\Phi^-\right) 
        - v^i(uv)\mukai{\dd^AC^-}{C^-}v^j\Phi^- \ ,  \\
\end{aligned}
\end{equation}
with $(\ee^{-C^-}D^\Ea\JJ_-)^{\Eb\Ec}$ given by complex conjugation and 
\begin{equation}
\begin{aligned}
   \big(\ee^{-C^-}D^{iA}\JJ_3\big)^j_{\hp{j}k}
      &= \tfrac{1}{4}v^i\dd^A\left[
            \kappa^{-1}\ii\mukai{\Phi^-}{\bar{\Phi}^-}
            (u^j\bar{u}_k+\bar{u}^ju_k)
         \right] 
         - \tfrac{1}{4}v^i\ii\kappa^{-1}
            \mukai{\Phi^-}{\bar{\Phi}^-}
            \\ & \qquad \times \mukai{\dd^AC^-}{C^-}\left[
               (uv)(v^j\bar{u}_k+\bar{u}^jv_k)
               + (\bar{u}v)(v^ju_k+u^jv_k)
           \right] \ , \\
   \big(\ee^{-C^-}D^{iA}\JJ_3\big)^B_{\hp{B}C}
      &= - \tfrac{1}{4}v^i\dd^A\left[ \kappa^{-1}
            \ii\mukai{\Phi^-}{\bar{\Phi}^-}(-\ii u\bar{u})
            \mathcal{J}^B{}_C \right] \ , \\
   \big(\ee^{-C^-}D^{iA}\JJ_3\big)^{j-}
      &= - \tfrac{1}{4}v^i\kappa^{-1}\ii\mukai{\Phi^-}{\bar{\Phi}^-}
            \\ & \qquad \times \Big[
            \left((\bar{u}v)u^j+(uv)\bar{u}^j\right)\dd^AC^-
            - \tfrac{1}{4}v^j(u\bar{u})
               \mathcal{J}_{BC}\Gamma^{BC}\dd^AC^-
            \Big] \ . 
\end{aligned}
\end{equation}
Using~\eqref{eq:connDK} and~\eqref{adjfund} to project on the
$\rep{56}$ component we then have
\begin{equation}
\begin{aligned}\label{Ppproject}
   \ee^{-C^-}\big(D\JJ_+\big)^{iA}
      &= v^i(uv) \left(
         \mukai{\dd^AC^-}{\Phi^-} +
         \mukai{\Phi^-}{\Gamma^A{}_B\dd^BC^-}
         \right) \ , \\
   \ee^{-C^-}\big(D\JJ_+\big)^+
      &= \dd\left[ (uv)\Phi^- \right] \ , 
\end{aligned}
\end{equation}
with again the complex conjugate expressions for $D\JJ^-$ and 
\begin{equation}
\begin{aligned}\label{P3project}
   \ee^{-C^-}\big(D\JJ_3\big)^{iA}
      &= \tfrac{1}{4}\dd^A\left[ \kappa^{-1}
            \ii\mukai{\Phi^-}{\bar{\Phi}^-}
            \left((\bar{u}v)u^i+(uv)\bar{u}^i\right)\right]
         \\ & \qquad \qquad 
         - \tfrac{1}{4}v^i\dd^B\left[ \kappa^{-1}
            \ii\mukai{\Phi^-}{\bar{\Phi}^-}(-\ii u\bar{u})
            \mathcal{J}^A{}_B \right]
         \\ & \qquad \qquad 
         - \tfrac{1}{2}\kappa^{-1}v^i(uv)(\bar{u}v)
            \ii\mukai{\Phi^-}{\bar{\Phi}^-}\mukai{\dd^AC^-}{C^-} \  , \\
   \ee^{-C^-}\big(D\JJ_3\big)^+ 
      &= - \tfrac{1}{2}\kappa^{-1}\ii\mukai{\Phi^-}{\bar{\Phi}^-}
         (uv)(\bar{u}v)\dd C^- \ . 
\end{aligned}
\end{equation} 
%
 


\end{document}